\documentclass[twocolumn]{aastex631}

\usepackage{xcolor}

\providecommand{\DIFdel}[1]{{}}

\usepackage[mathscr]{eucal} 
\usepackage{physics} 
\usepackage{multirow} 
\usepackage{tikz,amsmath,amssymb,amsthm} 
\usetikzlibrary{shapes,arrows.meta,positioning} 

\usepackage{subcaption} 

\let\tablenum\relax

\usepackage{siunitx} 
\usepackage{cancel} 
\usepackage{enumitem} 
\shorttitle{Escape of Secondary Atmospheres}

\begin{document}

\title{Novel Physics of Escaping Secondary Atmospheres May Shape the Cosmic Shoreline}

\author{Richard D. Chatterjee}
\author{Raymond T. Pierrehumbert}
\affiliation{Atmospheric, Oceanic and Planetary Physics, 
University of Oxford}
\email{richard.chatterjee@physics.ox.ac.uk}



\begin{abstract}
Recent James Webb Space Telescope observations of cool, rocky exoplanets reveal a probable lack of thick atmospheres, suggesting prevalent escape of the `secondary’ atmospheres formed after losing primordial hydrogen. Yet, simulations indicate that hydrodynamic escape of secondary atmospheres, composed of nitrogen and carbon dioxide, requires intense fluxes of ionizing radiation (XUV) to overcome the effects of high molecular weight and efficient line cooling. This transonic outflow of hot, ionized metals (not hydrogen) presents a novel astrophysical regime ripe for exploration. We introduce an analytic framework to determine which planets retain or lose their atmospheres, positioning them on either side of the cosmic shoreline. We model the radial structure of escaping atmospheres as polytropic expansions--- power-law relationships between density and temperature driven by local XUV heating. Our approach diagnoses line cooling with a three-level atom model and incorporates how ion-electron interactions reduce mean molecular weight. Crucially, hydrodynamic escape onsets for a threshold XUV flux dependent upon the atmosphere’s gravitational binding. Ensuing escape rates either scale linearly with XUV flux when weakly ionized (energy-limited) or are controlled by a collisional-radiative thermostat when strongly ionized. Thus, airlessness is determined by whether the XUV flux surpasses the critical threshold during the star’s active periods, accounting for expendable primordial hydrogen and revival by volcanism. We explore atmospheric escape from young Sun Mars and Earth, LHS 1140 b and c, and TRAPPIST-1 b. Our modeling characterizes the bottleneck of atmospheric loss on the occurrence of observable Earth-like habitats and offers analytic tools for future studies.
\end{abstract}


\section{Introduction} 
\label{sec: intro}
The circumstellar habitable zone is traditionally defined as where planets are cool enough to host liquid water \citep{Huang_1959}, but other conditions must be fulfilled to retain a suitable atmosphere. Making up more than 70\% of the stars in the Milky Way, the low-mass stars known as M dwarfs host at least one small planet on average \citep{Dressing_2015}. Low mass stars often do not spin down appreciably, meaning their ionizing luminosity can be one part in a thousand of their bolometric (total) luminosity for billions of years \citep{Fleming_2020}. For the population of planets companion to active stars, the open problem of what atmospheres will survive has risen in prominence due to the 2021 launch of the \textit{James Webb Space Telescope} with its capability to observe cool rocky planets in transmission and emission. Indeed, \textit{JWST} observations of the two closest-in planets of the now pre-eminent TRAPPIST-1 system \citep{GillonTrappist2017, Greene2023, Zieba2023} were consistent with bare rocks. If planets such as TRAPPIST-1 b and c are confirmed to be airless, is the most compelling explanation then a volatile-poor birth or catastrophic photoevaporation?

Meeting the conditions for the hydrodynamic escape of secondary atmospheres may divide the airless from the airy rocky exoplanet populations, metaphorically described as the Cosmic Shoreline \citep{Zahnle_2017}. Secondary atmospheres can form in the continuous transition from an escaping nebular atmosphere, evolve from an initial high-molecular-weight atmosphere, or be revived by volcanism from a bare rock \citep[e.g.,][]{KiteBarnett,KT2024}. We focus on modeling the retention of nitrogen-dominated atmospheres because this is crucial for assessing the potential for Earth-like habitats orbiting M dwarfs \citep{Lammer2019}. The physics governing the escape of CO$_{2}$-dominated atmospheres is similar but  differences arise due to infrared cooling lines and photochemistry \citep[e.g.,][]{Tian_2009, Van_Looveren_2024}. Though some insight can be carried over to the escape of steam-dominated atmospheres, the shift in dynamics due to the hydrogen ions in the upper atmosphere is outside the scope of the present study and we refer the reader to the literature devoted to this topic \citep[e.g.,][]{Johnstone_2020,Yoshida2022,Munoz2024}.

Because secondary atmospheres have a high molecular weight and efficient cooling by atomic lines, escaping them requires a large flux of stellar irradiation with a shorter wavelength than the hydrogen ionization edge: extreme ultraviolet (\qtyrange[range-units=single,range-phrase=-]{10}{92}{\nm}) and X-ray (\qtyrange[range-units=single,range-phrase=-]{0.1}{10}{\nm}). The strong and broad XUV photoabsorption cross-section finds the tenuous gas of the planet’s infrared photosphere optically thick and efficiently drives a temperature inversion from the hot electrons and atoms released. That is in contrast to the lower atmosphere, where heat from photoabsorption is efficiently reradiated as a blackbody to space, so it cannot contribute to escape. The work reported here constrains the threshold XUV fluxes for evolution to airlessness for the size range of rocky exoplanets.

\subsection{Hydrostatic versus Hydrodynamic}

\begin{figure}
\centering
\includegraphics[scale=0.37]{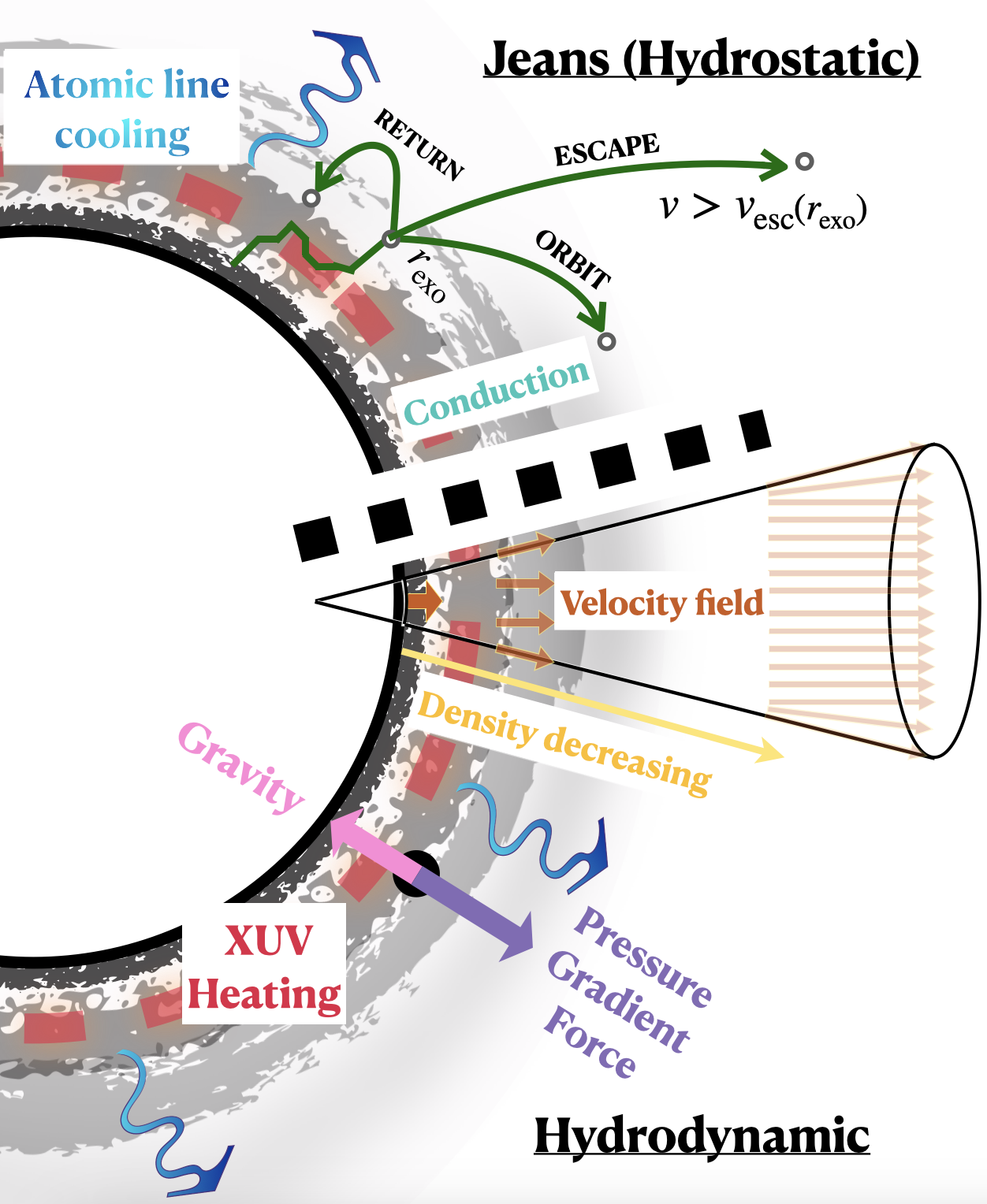}
\caption{Illustration comparing hydrodynamic and hydrostatic escape of a secondary atmosphere (not to scale). In hydrodynamic escape, the net heating from XUV photoabsorption, reduced by recombination and atomic line cooling, drives a pressure gradient force to overcome gravity--- accelerating the flow radially outwards down the atmosphere’s density gradient. Jeans escape occurs from the hydrostatic exobase where the atmosphere becomes collisionless. If the thermal velocity of a particle directed outwards exceeds the escape velocity $v_{\mathrm{esc}}$ at the exobase radius $r_{\mathrm{exo}}$, then it can ballistically escape on a hyperbolic trajectory. The balance of XUV heating, conduction, and line cooling determines the escape rate via the exobase temperature.}
\label{fig: hydroescape}
\end{figure}

The easier escape of hydrogen compared to nitrogen, or other metals, can be marked out through the two limits of thermal atmospheric loss: Jeans and hydrodynamic (Fig \ref{fig: hydroescape}). Jeans escape occurs at the exobase of a hydrostatic atmosphere, where collisions are infrequent, so the minor fraction of particles that are thermally endowed with the local escape velocity can be ballistically ejected to space \citep{jeans1921dynamical}. However, for significant depletion over millions of years, a mechanism of bulk escape is required. The atmosphere can flow into the interplanetary medium as a steady hydrodynamic wind, reaching supersonic speeds analogous to the solar wind predicted by \cite{Parker1958}. At the sonic point $r_{sc}$, the velocity must roughly equal half the local escape velocity $v_{esc}$. So, greater gravitational acceleration $g(r)$ or higher mean molecular mass $\mu$ of the flow requires higher atmospheric temperatures because $T_{sc}\propto g(r_{sc})r_{sc}\mu$. 

Atmospheric escape from Earth for an atomic nitrogen sonic flow requires a temperature of roughly \SI{15000}{\kelvin}, while for atomic hydrogen less than \SI{1000}{K} is sufficient. The complementary perspective is that a hydrostatic exobase cannot remain in equilibrium for a large rate of Jeans escape, which increases exponentially with temperature. When the sound speed reaches roughly half the escape velocity, the supply to the exobase shifts from diffusion to in-bulk. \cite{Tian2008} calculated that the exobase of a primordial hydrogen atmosphere on Earth would collapse at temperatures around \SI{500}{K} compared to \SI{7000}{K} for a composition of atomic nitrogen and oxygen.

So that the pressure gradient force can accelerate an outflow to sound speed, the flow must remain collisional--- the exobase should be above the sonic altitude. There is a corresponding threshold in XUV flux to meet this Knudsen onset, which we name after the dimensionless number for collisionality \citep{Knudsen1909}. This was called the transonic criterion by \cite{Johnson2013,Johnson2013b}, who explored the critical heating with molecular-kinetic simulations, with a focus on heating deposited in a single layer. However with increasing ionization, ion-neutral charge exchange and long-range Coulomb interactions make the Knudsen-onset less restrictive, leaving other conditions to be satisfied for transonic escape. 

Generally, the increase in escape with the higher temperatures generated by an increasing XUV flux limits the ionization fractions in an upper atmosphere. However, ions and electrons can still dominate if XUV heating can be efficiently radiated away. In which case, ambipolar motion will reduce the temperatures required for rapid escape: the Coulomb attraction of the electrons drags out the ions, effectively halving their mean molecular weight and so doubling their scale height \citep{Bauer&Lammer}. Earth's ambipolar electrostatic field was recently measured in-situ for the first time by the Endurance rocket \citep{EarthField}. The debated protective effects of planetary magnetic fields are beyond the scope of the present study.
 
\subsection{Radiative Cooling: Hydrogen versus Metal}
The photoevaporation of secondary atmospheres represents a novel astrophysical regime in that it is a `metal’ flow--- not dominated by hydrogen. The hydrodynamic escape of hydrogen atmospheres has been observed from hot Jupiters since \cite{Vidal2003, Vidal2004} and theoretical modelling was used to predict the bimodality in the radius distribution of sub-Neptunes \citep{Owen_2013, Lopez_2013, Fulton_2017}. For evaporation of hydrogen envelopes from close-in giant planets, XUV heating competes with cooling from electron-impact excitation of the HI\textsubscript{Ly$\alpha$} emission from \SI{10.2}{\electronvolt} above ground and by radiative recombination of ions with electrons \citep[e.g.,][]{Murray_Clay_2009}.

Groundbreaking simulations by \cite{Nakayama_2022} demonstrated that cooling via collisional excitation of the metastable states of N, C, and O—each with excitation energies around \qtyrange{1}{4}{\electronvolt}, making them more accessible than HI\textsubscript{Ly$\alpha$}—was sufficient to maintain temperatures low enough for an Earth-like atmosphere to remain hydrostatic under $1000\times$ current Solar XUV flux. Their inclusion of efficient atomic cooling revised the resistance of Earth's atmosphere to XUV-driven hydrodynamic escape by orders of magnitude \citep{Tian2008} and offered more favorable prospects for Earth-sized exoplanets orbiting M dwarfs retaining atmospheres. However, the present study reports that the assessment of Jeans escape by \cite{Nakayama_2022} to determine hydrostatic instability only partially accounts for the increasing ionization of the neutral exobase and neglects the eventual supplanting by the ion exobase.

We also note that self-consistency of the unexplored hydrodynamic equilibrium is not necessarily mutually exclusive with a self-consistent hydrostatic equilibrium as they occupy different regimes: ionization-recombination and ionization-advection balances, respectively. We will first explore predominantly neutral hydrodynamic escape and then introduce a new regime of global ion outflow controlled by a collisional-radiative thermostat. This thermostatic feedback, which can be inferred from the simulations of \cite{Nakayama_2022}, can be compared to that due to line cooling of minor-abundance metals in a photoionization-equilibrated hydrogen flow such as found in stellar winds \citep{DrewWind}, planetary nebula \citep{Seaton1960} and even galaxy clusters \citep{FerlandGalaxy}.  

\subsection{Analytic and Numerical Modeling}
\cite{Johnstone2019} time-stepped the discretised multispecies Navier-Stokes Equations with radiative transfer and photochemistry to solve for the steady-state escape of Earth’s atmosphere under the raised XUV output of the young Sun. This approach, henceforth referred to as a hydrocode, is the most comprehensive. However, its conclusions are still vulnerable to neglected physics, which in this case was the forbidden line cooling later found to be important by \cite{Nakayama_2022}, and the computational expense is prohibitive for broader studies. At the opposite end of complexity, there is the energy-limited parameterization, adapted from \cite{Watson1981}, which calculates the mass loss rate via the total XUV heating needed to accelerate atmospheric layers to escape velocity with some fixed efficiency. 

The work reported here uses a polytropic approximation for the atmospheric structure to bridge the modeling gap:
\begin{equation}
    \left ( \dfrac{\rho}{\rho_0} \right )^{\gamma -1} = \dfrac{T}{T_0}, \  0 < \gamma \leq \gamma_a,
\end{equation}
comparing the density $\rho$ and temperature $T$ to reference values $(\rho_0, T_0)$, for example, at the thermobase. By varying the polytropic index $\gamma$ between adiabatic expansion $\gamma_a$ (e.g. monatomic $\gamma_a=5/3$), Parker's isothermal wind $\gamma=1$, and thermal inversions $\gamma<1$, we unify a range of idealized planetary outflows. The polytropic equation of state is widely used but is best known for solutions to stellar structure \citep[e.g.,][]{Chandra}. We adopt and generalize results from application of polytropes to the transonic solar wind, including the exact algebraic solution to the dimensionless problem from \cite{Holzer1970} and extend approximations to the dimensional problem from \cite{lamers1999}. 

The other dimensionless parameter required to define polytropic solutions is the hydrodynamic escape parameter $\lambda_0$, which is the ratio of gravitational to thermal energy at the base of the flow. We show that $\lambda_0 \gg 1$ for high-molecular-weight atmospheres allows a priori specification of the sonic altitude and the base-to-sonic ratios of flow properties. \cite{Owen2012} use a polytropic model for hydrogen outflow from close-in hot Jupiters with $\lambda_0 \to \infty$ and $\gamma$ set by the slope of the photoionization equilibrium function. We show how $\gamma$ determines the profile of local polytropic heating to drive a given steepness of thermospheric inversion and relate $\gamma$ to the XUV flux given a fixed $\lambda_0$. \cite{Owen2012} extended their model to an altitude-dependent $\gamma$ via a pretabulated function; such a formulation may not be possible for metal outflows and is not explored in the present study. 

Exploring the transition from subsonic escape with direct simulation of particles \citep[DSMC,][]{bird1994}, \cite{Johnson2013,Johnson2013b} developed an analytic criterion for the minimum XUV heating to drive transonic outflow. We use the Knudsen onset to self-consistently specify a unique transonic $\gamma_{on}$ polytropic outflow launched from a base characterized by $\lambda_0$. We provide analytic expressions for the global thermomechanical efficiency and scale height at the sonic point. The polytropic index $\gamma_{on}$ yields a threshold XUV flux approximately constrained by a global energy limit and local heating at the sonic point. Forbidden atomic line emission finds the thermosphere optically thin, so we use a three-level atom model to analytically calculate cooling direct to space. When the cooling is found to exceed the polytropic heating in the Knudsen onset, conditions are not favorable for allowing predominantly neutral outflow. The limit of high XUV flux may instead drive global ion outflow under the collisional radiative thermostat.

\subsection{Structure of the Paper}
For the application of the polytropic framework and our other analytic models, we explore young Sun XUV-evaporation of nitrogen atmospheres from Earth and Mars, as well as prospects of atmospheric retention around M dwarfs in the cases of Earth-sized TRAPPIST-1 b and super-Earths LHS 1140 c and b. The remainder of the paper is organized as follows: 

Section \ref{sec: model} derives and characterizes the solution space for escaping planetary atmospheres under a polytropic equation of state that can be related to XUV heating. Section \ref{sec: onset} demonstrates how the Knudsen onset in collisionality determines a unique polytropic solution, establishing a threshold XUV flux required to sustain predominantly neutral transonic outflow of high molecular weight atmospheres. Section \ref{sec: nlte} introduces the three-level atom model for electron-impact excitation of forbidden lines, emphasizing its role in the energetics of the neutral Knudsen onset during transonic escape. Section \ref{sec: thermostat} explores whether atomic line cooling can protect a hydrostatic atmosphere from instability under a rising XUV flux producing more ions in the upper atmosphere as relevant to \cite{Nakayama_2022}. This leads us to introduce the regime of global ion outflow from super-Earths, where Coulomb collisions sustain pressure gradients within tenuous outflows, establishing alternative XUV-flux thresholds dictated by thermostat-limited escape. Section \ref{sec: discuss} integrates these findings to examine the broader landscape of high-molecular-weight atmospheric escape and its implications for the cosmic shoreline of rocky exoplanets. Our conclusions are summarized in Section \ref{sec: sum}.

\section{Polytropic model of atmospheric escape} 
\label{sec: model}
A compressible fluid in 1D requires four equations to solve for four unknowns: transport of mass, momentum, and energy, with an equation of state to calculate density, pressure, velocity, and temperature. The polytropic relation (Eqn \ref{eqn: polytropic-expansion}), in addition to the ideal gas law, allows us to avoid integrating the momentum equation. However, the momentum equation is still key to understanding the flow solution. In this section, we build on previous work to derive an analytic solution space for the structure of escaping atmospheres.  

\subsection{Character of XUV-driven Transonic Expansion into Space}
A planetary atmosphere, driven by XUV heating, expands outward from the thermobase, meets the stellar wind, and matches its dynamic pressure, yielding a subsonic breeze with a mass loss rate proportional to the `inlet-outlet' pressure difference. Continuously reducing the pressure imposed by the stellar wind allows further outward acceleration of the atmosphere driven by the pressure gradient until the flow becomes supersonic at a critical pressure. Once the flow is transonic, the mass loss rate has reached a maximum independent of the upper boundary conditions: any forcing downstream of the sonic altitude cannot influence the upstream flow \citep{pierrehumbert_2010}. A subsonic breeze transporting gas up to the exobase is also possible, yielding ballistic escape from a drifting Maxwellian \citep{Volkov2011b}, but we set this case aside, noting that it would still respect the upper bound of the transonic rate.

\begin{figure}
\plotone{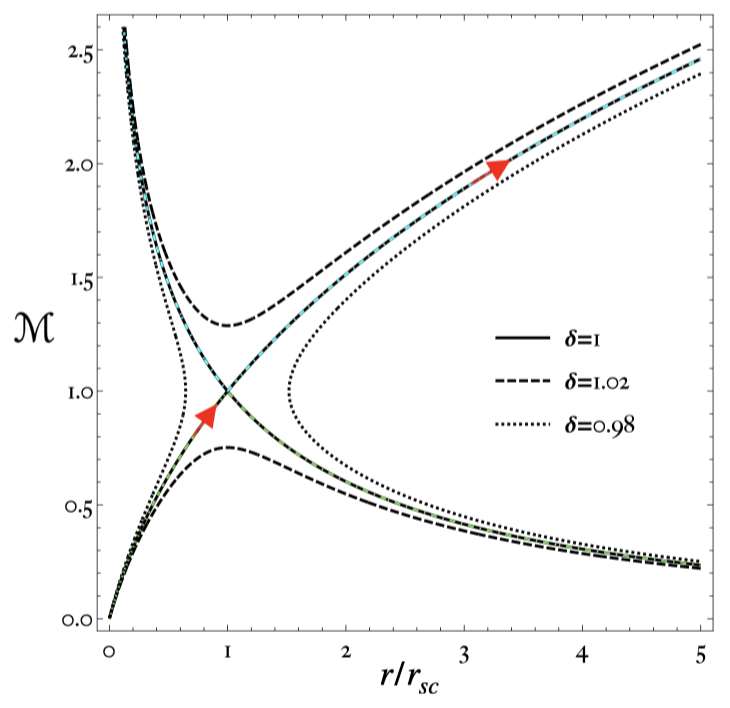}
\caption{Contour plot of Mach number $\mathscr{M}$ against dimensionless radius $r/r_{sc}$ for a diatomic adiabatic expansion ($\gamma_a=1.4$) example solutions: transonic ($\delta =1$), subsonic/supersonic ($\delta=1.02$), and double-valued ($\delta=0.98$). The transonic escape solution is highlighted with red arrows. The transonic contours' upper and lower branches are highlighted with a dot-dashed line of cyan and green, respectively. $\delta$ relates to the entropy differences between different types of solutions (see Eqn \ref{eqn: isenexp}). \label{fig: contours}}
\end{figure}

The spherically symmetric steady-state compressible Euler Equations admit a range of flow solutions in a gravitational field, illustrated in the Mach-radius phase plane in Figure \ref{fig: contours}. The integral curves form a saddle geometry centered around the sonic point: $\mathscr{M}=1$, $u_{sc} = \sqrt{fRT_{sc}}, \ r=r_{sc}$, where $f$ is an order unity dimensionless prefactor, and R is the specific gas constant. One way to interpret the geometry is to note that a flow can evolve over time to occupy different parts of the solution curves for accretion and escape through a hysteresis loop \citep{Velli}.

The mass flux of the outflow must be radially uniform, and the momentum equation can then be written as a compressibility-weighted acceleration outwards, provided by a pressure-gradient force split into an adiabatic and a diabatic component:  
\begin{equation}
    (u^2 - c_{a}^2) \frac{1}{u} \dv{u}{r} = \frac{2 c_{a}^2}{r^2} \left (r - \frac{GM}{2c_{a}^2} \right) - (\gamma_a - 1) 4 \pi r^2 q_{v},
    \label{eqn: transonic rule}
\end{equation}   
with radius $r$, radial velocity $u$, adiabatic sound speed $c_{a} =\sqrt{\gamma_{a}RT}$, net volumetric heating rate  $q_{v}$, density $\rho$, and planet mass $M$. 

For volumetric heating prescribed as a function of radius only, the velocity gradient is unbounded where the flow velocity equals the adiabatic sound speed unless the radius satisfies $r_{sc}=GM/2u_{sc, a}^2$. However, $q_{v}(r)$ is not representative of planetary atmospheres: if the flow is perturbed, the density changes, which would displace the distribution of stellar heating such that heating is implicitly also a function of the velocity gradient.
To find the sonic point for $q_{v}(r, \dv{u}{r})$ in general, if one exists at all, the energy and momentum equations can be written as an autonomous system of decoupled first-order ODEs \citep{Bauer2021}. The central reduction of the present study is to an approximately polytropic atmospheric structure, given by 
\begin{equation}
   \left (\dfrac{\rho}{\rho_0} \right)^{\gamma -1} = \dfrac{T}{T_0}, \  u_{sc}= \sqrt{\gamma R T_{sc}} \ \& \ r_{sc}= \frac{GM}{2 u_{sc}^2} .\label{eqn: polytropic-expansion}
\end{equation}
This equation generalises from the $\gamma =\gamma_a$ adiabatic expansion to a polytropic expansion where $ 0 < \gamma < \gamma_a$, corresponding to implicit inclusion of distributed heating given by
\begin{equation}
    q_{\gamma} \cdot 4 \pi r^2 \rho  = \Phi_{\mathrm{hyd}} \cdot \left( \frac{\gamma_a}{\gamma_a - 1} + \frac{\gamma}{1 - \gamma} \right) R \dv{T}{r}   ,\label{eqn: poly-heat}
\end{equation}
where $\Phi_{\mathrm{hyd}}$ is the mass loss rate and $q_{\gamma}$ is the polytropic model’s implicit specific heating per unit mass.
The polytropic solution $(\Gamma=\gamma)$ will be consistent with a diabatic $(\gamma_a, q)$ solution at each altitude: 
\begin{equation}
\Phi_{\mathrm{hyd}}  \cdot \dv{r} \left( \frac{u^2}{2} + \frac{c_{\Gamma}^{2}}{\Gamma -1}  -\frac{GM}{r} \right) = \begin{cases}  4 \pi r^2 \rho q & : \Gamma = \gamma_a \\ 0 & : \Gamma = \gamma, \end{cases} \label{eqn: comparison}
\end{equation}
if $q=q_{\gamma}$.
The differently framed flows have the same velocity and gravitational potential but different internal energies. The physical interpretation of Eqns (\ref{eqn: poly-heat} \& \ref{eqn: comparison}) is that the total heating in each local spherical shell drives the transport of specific internal energy at each altitude. The shift of the saddle point from the adiabatic $u/c_{a}$ to the polytropic Mach number $u/c_{\gamma}$ is due to a regularising perturbation from the imposition of polytropic heating, as the temperature becomes implicitly dependent on the velocity gradient through the implicit heating. This shift in the sonic point is comparable to that from conduction regularization \citep[e.g.,][]{Hong2014}. Note that the classic analytic solution of the isothermal Parker Wind also requires implicit heating to accelerate the flow, which the XUV section of stellar irradiation often provides. 

A hydrodynamic upper atmosphere, not limited by the supply of the escaping component from the lower atmosphere, will often escape limited by the energy input \citep[e.g.,][]{Watson1981}. The absorbed stellar XUV lifts gas out of the planet's gravitational well with an efficiency $\eta$: 
\begin{equation}
\Phi_{\mathrm{hyd}} \frac{GM_{p}}{r_0}  = \eta \cdot \pi r_{abs}^2 \frac{\mathscr{L}_{\scriptscriptstyle \mathrm{XUV}}}{4 \pi a^2}, \label{eqn: E-L}
\end{equation}
where the planet's solid surface radius is $r_0$, orbiting with semi-major axis $a$; $\mathscr{L}_{\scriptscriptstyle \mathrm{XUV}}$ is the star's XUV luminosity with a peak absorption at planetary radius $r_{abs}$. For what intensity of XUV irradiation is the energy-limited parametrization a good approximation? And for which atmospheres may cooling outweigh heating so that application of the energy-limit becomes unclear? The following analysis provides answers to these questions through the Knudsen onset and the polytropic profiles from which to diagnose heating and cooling. For hydrogen escape, an alternative prescription for advancing beyond the energy-limited parametrization used a grid of hydrocode models to construct fits for mass loss rates \citep{Kubyshkina_2018}. 

The present study breaks down the efficiency $\eta$ into a product of the thermomechanical, photoreaction and cooling efficiencies:
\begin{subequations}
\begin{eqnarray}
    \eta &=& \eta_{\gamma} \eta_{\mathrm{pr} } \eta_{\mathrm{c}}, \\ \label{eqn: efficiency}
    \eta_{\gamma} &\approx& \frac{2 \gamma r_{sc} }{r_0} \left ( \frac{\gamma_a}{\gamma_a -1 } + \frac{\gamma}{1 - \gamma}\right)^{-1} \label{eqn: TM}
\end{eqnarray} 
\end{subequations}
The thermo-mechanical $\eta_{\gamma}$ efficiency of polytropic escape is derived by integrating the implicit heating up to the sonic point (Eqn \ref{eqn: poly-heat}) and comparing to the heating needed to lift the mass out of the gravitational well. The thermal energy at the base is neglected, so Equation (\ref{eqn: TM}) is approximate, but an exact comparison as a function of the boundary conditions is straightforward. The photoreaction efficiency accounts for the excess energies of photoionization and photodissociation. The cooling efficiency represents how much of the available XUV-heating is converted into line cooling from the thermosphere, which will be explored in Section \ref{subsec: LC}.

\subsection{Polytropic Approximation to Atmospheric Structure and Escape}

\begin{figure*}
\centering
\includegraphics[scale=0.4]{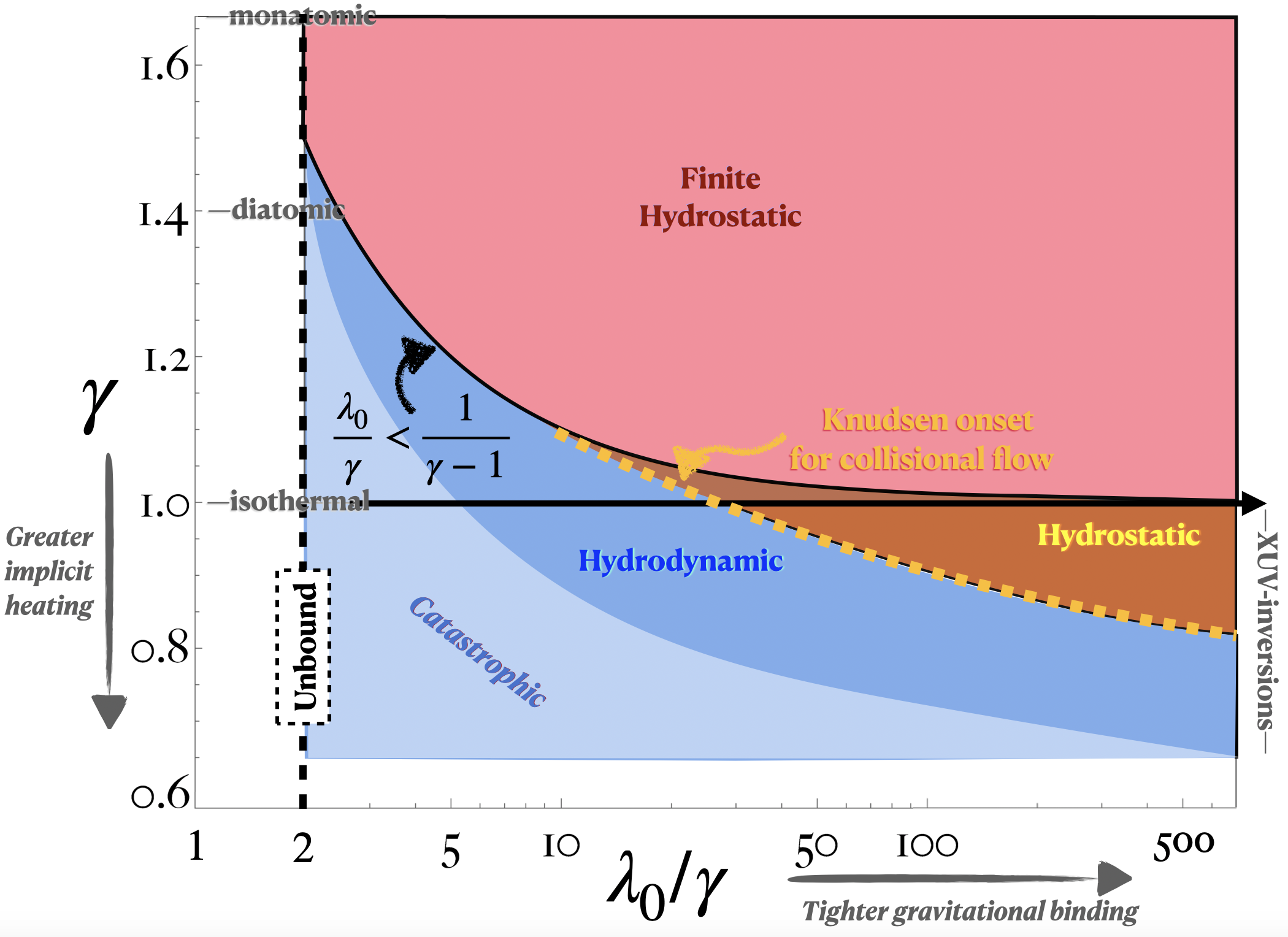}
\caption{Steady-state polytropic solutions space for vertical atmospheric equilibrium. For $\frac{\lambda_0}{\gamma} > \frac{1}{\gamma-1}$, the equilibrium is hydrostatic with density reaching zero at a finite height (pink). Implicit heating throughout the atmosphere reduces $\gamma$ from an adiabat $\gamma_a$. For $2<\frac{\lambda_0}{\gamma} <\frac{1}{\gamma-1}$, the thermal energy at the base in addition to the implicit heating is large enough to drive an accelerating transonic flow in principle (blue hydrodynamic region). The supersonic at the base constraint $\lambda_0=2$ is represented by the dashed vertical line and labelled `unbound'. The blue region of consistent steady states does not extend all the way to this line, however, because the flow launch should have $\mathscr{M} \ll 1$, otherwise the escape is `catastrophic' (illustrated in light blue). On failure of the hydrodynamic assumption at the sonic point, XUV-driven inversions ($\gamma < 1$) give way to hydrostatic atmospheres that would be infinite except for the formation of an exobase (brown region). The yellow line illustrates the Knudsen onset for collisional flow given by $\mathrm{Kn}_{sc}^{\mathscr{N}} \sim 1$ (see Section \ref{sec: onset}). \label{fig: SolSpace}}
\end{figure*}

Polytropic atmospheric structure spans a range of idealized hydrostatic and hydrodynamic vertical equilibria. \citet{Holzer1970}, following on from the work of Parker, provided an algebraic solution to the dimensionless problem for the polytropic solar wind. Our work extends the calculation to solve the dimensional problem, folding it into the framework for high-molecular-weight transonic escape of planetary atmospheres.    
Conservation of mass and the Bernoulli equation for energy link the sonic flow properties to the rest of the flow as:
\begin{subequations}
\begin{eqnarray}
\rho u r^2 & = &\rho_{sc} u_{sc} r_{sc}^2, \label{eqn: mass flux} \\ 
\frac{u^2}{2}+\frac{c^2}{\gamma-1}-\frac{G M}{r} & = &\left(\frac{5-3 \gamma}{\gamma-1}\right) \frac{u_{sc}^2}{2}, \label{eqn: bernoulli}
\end{eqnarray}
\end{subequations}
where we have simplified the energy at the sonic point using Equation (\ref{eqn: polytropic-expansion}). This pair of equations can be expressed in terms of two unknowns, the Mach number $\mathscr{M} = u/c_{\gamma}$ and dimensionless radius $\xi = r/r_{sc}$, yielding
\begin{subequations}
\label{eqn: constitutive}
\begin{eqnarray}
\left ( \frac{c^2}{u_{sc}^{2}} \right)^\frac{\gamma+1}{\gamma-1} \mathscr{M}^2 \xi^{4} &=&1, \label{eqn: nonDmass}
\\
\frac{c^2}{u_{sc}^2}\ \left(\frac{\mathscr{M}^2}{2} +\frac{1}{\gamma -1}\right)&=&\frac{1}{2} \left(\frac{5-3 \gamma}{\gamma-1} \right)+ \frac{2}{\xi}. \label{eqn: nonDenergy}
\end{eqnarray}
\end{subequations} 
To find the implicit algebraic solution, $\frac{c^2}{u_c^2}$ can be eliminated from Equation (\ref{eqn: nonDenergy}) by substitution of Equation (\ref{eqn: nonDmass}). The governing equation
\begin{equation}
2 \delta \left(\frac{1}{\gamma-1}+\frac{\mathscr{M}^2}{2}\right)\left(\mathscr{M}^2\right)^{ -\frac{\gamma-1}{\gamma+1}}=\left(\frac{5-3 \gamma}{\gamma-1}+\frac{4}{\xi}\right) \xi^{ 4 \left(\frac{ \gamma-1}{\gamma+1}\right)} \label{eqn: isenexp}
\end{equation}
can be found numerically as a contour, where the transonic outflow is selected from amongst accretion and multivalued solutions (Fig \ref{fig: contours}) via specifying a constant of integration $\delta = 1$ \citep{Holzer1970}. Related to entropy, we have reinserted $\delta$ so that Equation (\ref{eqn: isenexp}) is in its most general form. The form of solution mirrors the isothermal Parker Wind, except that most scientific computational libraries have a pre-tabulated function for the Parker contours, called the Lambert W-Function \citep{Cranmer2004}. Equation (\ref{eqn: isenexp}) also recovers the isothermal solution in the limit $\gamma \to 1$ with the Bernoulli definition of the exponential; see discussion of Equation (\ref{eqn: kin_poly}). 

The above solution for the Mach profile is the seed for the solution to all other profiles. To calculate the dimensionless temperature, conservation of mass (Eqn \ref{eqn: nonDmass}) can be used directly. The temperature profile can be fed into the polytropic relationship (Eqn \ref{eqn: polytropic-expansion}) to determine the dimensionless density profile. Lastly, a combination of the temperature and Mach number profiles yields the dimensionless velocity profile via energy conservation (Eqn \ref{eqn: nonDenergy}).

The hydrodynamic escape parameter is defined as the ratio of gravitational energy to thermal energy at the base of the flow \citep[e.g.,][]{Watson1981}:
\begin{equation}
    \lambda (r_0)= \lambda_0 = \frac{GM}{R T_0 r_0} \label{eqn: HEP}
\end{equation}
and is the other dimensionless parameter for the polytropic model. We will refer to an escape temperature analogous to the concept from protoplanetary disc photoevaporation \citep[e.g.,][]{Owen2012}. It is defined as the temperature necessary at some altitude for the isothermal sound speed to equal half the escape velocity, or equivalently, $\lambda(r)=2$: 
\begin{equation}
    T_{\mathrm{esc}}(r) = \frac{\mu }{4  k_{B}} v_{\mathrm{esc}}(r)^2 . \label{eqn: Tesc}
\end{equation}
At the escape temperature, the sonic point  of an isothermal hydrodynamic flow would be reached (Eqn \ref{eqn: polytropic-expansion}), while if the top of a hydrostatic atmosphere were at this temperature it would collapse into a blow-off state \citep{Opik-blowoff}. Moreover, if the temperature at the exobase is a half to a third of the escape temperature, a hydrostatic equilibrium would be unstable and expand hydrodynamically \citep{Tian2008}.  

Although these profiles can hypothetically be calculated for a planetary atmosphere with any $(\gamma, \lambda_0)$, for a well-behaved steady-state solution, the planetary wind should launch highly subsonic. Thus, the $\mathscr{M}^2$ term in energy conservation can be neglected to provide a simple approximation for the sonic radius, where
\begin{equation}
    \xi_0 \approx \frac{\frac{4 \gamma}{\lambda_0}-4(\gamma -1)}{5-3\gamma} =  \xi_{0}^{\mathrm{itr},1}, \label{eqn: itr}
\end{equation}
specifying the dimensional profile and collapsing the degeneracy in the sonic point through a single iteration in the first approximation $\xi_{0}^{\mathrm{itr},1}$ \citep{lamers1999}. In the context of flows which start with appreciable Mach numbers, one can find the $\mathcal{M}^2{ }_{\mathrm{itr}, 1}$ via the Equations (\ref{eqn: itr} \& \ref{eqn: isenexp}), to find the second approximation, given by
\begin{equation}
\xi_0^{\mathrm{itr}, 2} = \frac{\frac{4 \gamma}{\lambda_0}\left(\frac{1}{2} \mathcal{M}^2{ }_{\mathrm{itr}, 1}+\frac{1}{\gamma-1}\right)-4}{\frac{5-3 \gamma}{\gamma-1}}.
\end{equation} 
This allows full specification of, for example, adiabatic diatomic expansion, usually after only one iteration. The equations for the sonic point radius can be compared to  Equation (13) from \cite{Owen2012}, which assumed $\gamma < 1, \lambda_0 \to \infty$.

Neatly, the approximate mass flux can be found immediately using the first iteration (Eqn \ref{eqn: itr}) - without needing to compute the contour. The mass loss rate, given by $4 \pi \rho u r^2$ and expressed in terms of $\gamma, \lambda_0$ and the lower boundary, becomes 
\begin{multline}
    \Phi_{\mathrm{hyd}} = 4 \pi \rho_0 r_0^2 \left(\frac{v_{\mathrm{esc}}(r_0)}{2}\right)  \left(\frac{\lambda_0}{2 \gamma}\right)^{1/(\gamma -1)} \\  \times \left(\frac{4\gamma/\lambda_0 - 4(\gamma-1)}{5-3\gamma}\right)^{\dfrac{5-3\gamma}{2(\gamma-1)}},  \label{eqn: mass loss}
\end{multline}
which has also been applied to stellar winds \citep{lamers1999}.
Lastly, Equation (\ref{eqn: itr}) allows simple determination of all sonic-to-base ratios for flow launching highly subsonic: 
\begin{subequations}
\begin{eqnarray}
\frac{T_{sc}}{T_0} & = & \frac{\lambda_0 \xi_0}{2\gamma}, \label{eqn: Tsonic} \\ 
\frac{\rho_{sc}}{\rho_0} & = & \left (\frac{T_{sc}}{T_0} \right)^{1/(\gamma-1)} \label{eqn: rho-sc}
\\ 
\frac{u_{sc}}{u_0} & = & \frac{\rho_{0}}{\rho_{sc}}  \xi_0^2 .
\label{eqn: u-sc}
\end{eqnarray}
\end{subequations}
For tightly bound atmospheres, where \(\lambda_0 \gg 1\), and \(\gamma < 1\),  variations in \(1/\xi_0\) are of order unity and the term with the strongest dependence on \(\gamma\) arises from the density (see Eqn.~\ref{eqn: rho-sc}). On varying \(\gamma\), the mass loss rate and the fuel required from XUV heating, thus scale as \(\propto \lambda_0^{1/(\gamma - 1)}\). The dimensions of the mass loss and XUV flux are $\rho_0 r_0^2 v_{esc}$ and $\rho_0 v_{esc}^3$ respectively (Eqns \ref{eqn: mass loss} \& \ref{eqn: fluxon}), though solutions can be constructed so as to be insensitive to choice of $\rho_0$ (Sections \ref{subsec: LB} \& \ref{subsec: Mars-onset}). The broader landscape of \(\gamma\) is examined in more detail in Figures~\ref{fig: SolSpace} and~\ref{fig: conceit}.
\subsection{Polytropic Solution Space}

Figure \ref{fig: SolSpace} presents the solution space of vertical equilibrium for polytropic atmospheres. We devote significant discussion to the space of regions in $(\gamma,\lambda_0)$ so that Figure \ref{fig: SolSpace} can be used as a touchpoint to understand idealized solutions to escape. It is instructive to follow the thermodynamic expansion of a parcel lifted through the atmosphere.

For $\lambda_0/\gamma > 1/(\gamma -1)$, the sum of the thermal energy at the base with the integrated heating is insufficient to give the parcel enough energy to escape the planet’s gravitational well so that the atmosphere will be hydrostatic. An air parcel will cool as it rises, and its density will tend to zero on reaching a finite terminus of the atmosphere. The cooling will be less for $\gamma \to 1$, where $\gamma < \gamma_a$ entails implicit heating throughout the atmosphere. The pink region of Figure \ref{fig: SolSpace} thus marks where only a finite hydrostatic atmosphere is consistent.

Moving leftwards, the blue region $\lambda_0/\gamma < 1/(\gamma -1)$ in Figure \ref{fig: SolSpace} marks where transonic escape solutions exist, and finite hydrostatic atmospheres are inconsistent. There are no consistent hydrodynamic escape solutions for $\gamma > 3/2$, as such flows would be decelerating from the base and only preserve uniform mass transport outwards because the spherical surface area increases with the radius. The $\gamma =1.4$ diatomic adiabatic expansion permits an accelerating transonic solution with an unrealistic velocity profile. The near field limit $\xi \ll 1$ of Equation (\ref{eqn: isenexp}) combined with Equation (\ref{eqn: nonDmass}) yields $u_0/u_{sc} \approx 1.3 \sqrt{\mathscr{M}_0}=\sqrt{2\mathscr{\xi}_0}$, so that even as far out as $r_{sc} = 40 r_0$, which allows $\mathscr{M}_0 = 1/20$, the sonic velocity is only $3 \times$ the base velocity in the diatomic case. Then, the atmosphere above the base is lost on the timescale of days.

Reducing $\gamma$ down towards 1 through the implicit imposition of heating allows the flow to be driven from a lower temperature base and for acceleration to sonic velocity orders of magnitude greater than base velocity, yielding loss timescales for the thermosphere on a timescale of thousands of years or more. Replenishment of the thermosphere by the lower atmosphere enables a quasi-steady-state hydrodynamic escape solution that evolves over a loss timescale of about a million years, with atmospheric pressure either decreasing or being maintained by outgassing.

In an isothermal expansion $\gamma=1$, the gas does work but heat flows in to keep the temperature the same. 
For $\gamma < 1$, a rising parcel heats as it expands up a thermospheric inversion due to the implicit heating outweighing adiabatic cooling. The model mass loss then increases linearly with the XUV flux, apart from a small variation in the thermomechanical efficiency, consistent with the energy limit.

Two exceptional regions show where mass loss is catastrophically fast in light blue and where the hypothetical flow becomes collisionless before reaching sound speed in brown. The catastrophic region begins where an atmosphere would be unbound at $\lambda_0 \leq 2$. Even for higher $\lambda_0$, increasing XUV flux reduces $\gamma$ so that sound speed is reached in a dense layer (Eqn \ref{eqn: rho-sc}), yielding too-rapid mass loss for a valid steady state of a planetary atmosphere. In the limit of large $\lambda_0$, a hydrodynamic atmosphere would find the exobase below the hypothetical sonic altitude. The Knudsen onset (yellow-line), which will be discussed in detail in the following section, entails a minimum steepness of polytropic inversion $\gamma_{\scriptscriptstyle \mathrm{on}}$, which must be generated by a threshold XUV flux to drive transonic hydrodynamic escape.  

\subsection{Lower boundary dependence}
\label{subsec: LB}
The role of lower boundary conditions varies with the polytropic index. For adiabatic thermally driven escape, any gas at the bottom of the potential well has enough thermal energy to escape, so the mass loss rate is set by the base density. Similarly, isothermal planetary winds are set by lower-boundary assumptions with large uncertainties and degenerate scenarios arise; see core-powered mass loss \citep[e.g.,][]{Tang_2024}. Steep inversions ($\gamma<1$) are the only analytic solutions that are consistent with the dynamics of the energy limit. The densities in the flow (base or sonic) are restricted to yield a mass loss less than the XUV heating available. We adopt a temperature close to equilibrium $T_{eq}$ and thermobase-like densities for the lower boundary of secondary atmospheres. We note that a spectrum richer in shorter wavelengths will launch the flow from deeper into the atmosphere, all else being equal. As will be discussed in Section \ref{subsec: Mars-onset}, when the base density is varied by an order of magnitude, the profiles are only moderately sensitive, and the mass loss rate is quite insensitive, remaining within the error already anticipated by the analytic approximation.
\section{XUV-flux Threshold for Transonic Escape} 
\label{sec: onset}
In this section, we will calculate whether there is an onset of transonic escape of hypothetical nitrogen atmospheres from Earth and Mars on ramping up XUV insolation to that experienced in the early pre-main sequence. Given best estimates of lower boundary conditions, we derive how imposing a Knudsen number equal to one at the sonic point specifies a unique polytropic solution to atmospheric escape. The corresponding threshold in XUV flux to drive the transonic outflow defines a conservative regime of applicability for energy-limited parametrization. Our ansatz is predominantly neutral outflow with XUV heating fueling advection. In Sections \ref{sec: nlte} \& \ref{sec: thermostat} we will relax these assumptions.
\subsection{Knudsen Onset}
\label{subsec: onset}
A Knudsen number equal to one at the sonic point will specify the escape of a polytropic atmosphere sustainable for an XUV flux that defines an energetic threshold for transonic escape. The Knudsen number \citep{Knudsen1909} describes the collisionality of a flow, which is calculated for an atmosphere as the ratio of the mean free path to the local scale height of the atmosphere. We derive the scale height using L'hopital's rule for the sonic velocity gradient \citep{lamers1999}, then substitute into the radial derivative of equation (\ref{eqn: mass flux}), to yield 
\begin{equation}
\left [\frac{1}{\rho} \left (\dv{\rho}{r} \right)_{sc} \right]^{-1} = - \frac{(1+\gamma) r_{sc}}{4+\sqrt{2} \sqrt{5-3 \gamma}}. \label{eqn: scale-height} 
\end{equation}
The mean free path is the average distance a particle will travel before its next collision. It can be calculated as the reciprocal of the product of the collision cross-section $\sigma_{\scriptscriptstyle C}$ and the number density up to a factor of order unity. Thus, the Knudsen number in the polytropic model can be expressed as 
\begin{equation}
\mathrm{Kn}_{sc}=\frac{4+\sqrt{2} \sqrt{5-3 \gamma}}{\sqrt{2}(1+\gamma) \sigma_{\mathrm{ \scriptscriptstyle C}} \cdot n\left(r_{s c}\right) \cdot r_{sc}} \label{eqn: knudsen} 
\end{equation}
The model exobase separates a collisional thermosphere below ($\mathrm{Kn} < 1$) from the collisionless exophere above ($\mathrm{Kn} > 1$). We compare our criterion for transonic escape with the numerical and analytic calculations of \cite{Johnson2013} in Section \ref{subsec: Johnson}. We note that \cite{Owen2012} found ballistic escape from some more massive hot Jupiters on the basis of non-existence of a collisional sonic point but did not explicitly relate this to their polytropic model or XUV flux. Furthermore, \citet{2016Owen} discuss how Kn$_{sc}$ is smaller for puffier hydrogen atmospheres on terrestrial planets, which implies a greater mass loss rate. 

The Knudsen onset for transonic escape should be taken with two caveats:
\begin{enumerate}[label=\Roman*.,noitemsep,topsep=0pt,parsep=0pt,partopsep=0pt]
\item Collisionality increases with greater ionization due to the greater cross-sections of ion-atom charge exchange and high-frequency atom-electron collisions. To establish a general method to calculate the onset of transonic escape, we choose to work with the neutral onset $\mathrm{Kn}_{sc}^{\mathscr{N}} \sim 1$, using the neutral collision cross-section $\sigma_{ \scriptscriptstyle \mathrm{C}}^{\mathscr{N}}$; reasonable when advection-dominated and weakly ionized.
\item In a bulk outflow, the nominal exobase forming before the hypothetical sonic point is consistent with the upper boundary fulfilling the flux constraints of a drifting-Maxwellian Jeans escape \citep{Volkov2011b, Erwin2013}. The energy limit might apply for  $\mathrm{Kn}_{sc} \in [1, 3]$, or extend even further \citep{Johnson2013}, but characterizing the non-linear transition requires both DSMC and hydrocode modeling, so it is beyond the scope of the present study.
\end{enumerate}
Thus, for characterizing rapid mass loss, the neutral onset is highly conservative.

The crux of the polytropic framework is to estimate the threshold XUV flux needed to drive (Eqn \ref{eqn: poly-heat}) transonic hydrodynamic escape for atmospheres of varying gravitational binding and composition. Temperatures and ion production should not be so extreme that line-cooling chokes off the flow, which will be discussed in Section \ref{sec: nlte}. The onset XUV flux is determined by the mass loss rate (Eqn \ref{eqn: mass loss}) for the polytropic atmosphere $\gamma_{\mathrm{on}}$ where $\mathrm{Kn}_{sc}=1$ :
\begin{equation}
    F_{\scriptscriptstyle \mathrm{XUV}}^{\mathrm{on}} = \frac{\Phi_{\mathrm{hyd}}(\gamma_{\mathrm{on}};\lambda_0, r_0, n_{0}, \mu, \sigma_{\scriptscriptstyle C}^{\mathscr{N}}) v_{\mathrm{esc}, 0}^{\ 2}}{2 \pi r_{abs}^2 \eta_{\gamma} \eta_{\mathrm{pr} } \eta_{\mathrm{c}} }, \label{eqn: fluxon}
\end{equation}
mediated by the thermomechanical efficiency $\eta_{\gamma}$ (Eqn \ref{eqn: efficiency}), the photoreaction efficiency $\eta_{\mathrm{pr}}$ and the cooling efficiency $\eta_{\mathrm{c}}$. The greater the collisionality required at the sonic point, the smaller $\gamma_{\mathrm{on}}$ becomes, with the XUV flux rising proportionally with the mass loss rate beyond the onset values.

\begin{widetext}
To find $\gamma$ as a function of $\mathrm{Kn}_{sc}$, we substitute the sonic-to-base density ratio (Eqn \ref{eqn: rho-sc}) into Equation (\ref{eqn: knudsen}):
\begin{equation}
    \frac{ (1+\gamma) (5-3\gamma) \lambda_0}{4(4+\sqrt{2} \sqrt{5-3 \gamma})( \gamma-\lambda_0\gamma + \lambda_0)} \cdot \left ( \frac{(5-3 \gamma) \gamma}{2 \gamma + 2\lambda_0  (1-\gamma)} \right )^{1/(1-\gamma)} - \left (\sqrt{2} \mathrm{Kn}_{sc}   \sigma_{ \scriptscriptstyle \mathrm{C}}  n_{0} r_0 \right )^{-1} = 0 \label{eqn: kin_poly}, 
\end{equation}
which defines a dummy function $\psi$, such that the neutral onset is given by $\gamma_{\mathrm{on}}= \psi \left(\mathrm{Kn}_{sc}=1; \lambda_0, r_0, n_{0}, \sigma_{\scriptscriptstyle C}^{\mathscr{N}} \right)$. The parameters after the semicolon in $\psi$ are specified as constants for a particular planetary atmosphere in a given regime. We can use the original Bernoulli definition of the exponential $\lim_{j \to \infty} \left (1 + \frac{x}{j} \right)^{j} = \mathrm{e}^x$, to recover the isothermal limit,  of Equation (\ref{eqn: kin_poly}) with  $j= 1/(1-\gamma)$
\begin{equation}
    \lim_{j \to \infty} \left ( \frac{1 + \frac{3}{2j}}{1 + \frac{\lambda_0}{j}} \right )^{j} \propto \frac{\lambda_0}{6} \cdot \mathrm{e}^{3/2 - \lambda_0} = \left (\sqrt{2} \mathrm{Kn}_{sc} \sigma_{ \scriptscriptstyle \mathrm{C}} \cdot n(r_0) \cdot r_0 \right )^{-1}, \label{eqn: kin_iso}
\end{equation}
where the derivation requires Taylor expansion of the denominator. We note that the isothermal solution restricts $\mathrm{Kn}_{sc}$ to be specified by the lower boundary conditions. In the polytropic framework, given the best-estimate base density and neutral cross sections, $\gamma_{\mathrm{on}}$ is found implicitly through Equation (\ref{eqn: kin_poly}) and selects the contour for the atmospheric structure (Eqn \ref{eqn: isenexp}). The corresponding flux is given by Equation (\ref{eqn: fluxon}), with a further estimate needed for the absorption radius. 
\end{widetext}

The underlying equations (\ref{eqn: fluxon}–\ref{eqn: kin_poly}) should be approximately consistent with local heating rates, especially at the sonic point. Assuming the flow at the sonic point is optically thin to incoming XUV, a physical estimate of XUV heating per unit mass at the sonic point is 
\begin{equation}
\overline{q_{sc}} \approx a_{\theta} \sum_{\lambda}
\eta_{pr}(\lambda) F_{\mathrm{xuv}}(\lambda) \sigma_{\lambda}/ m, \label{eqn: heat_thin}
\end{equation}
where $m$ is the atomic mass, $\sigma_{\lambda}$ and $\eta_{pr}(\lambda)$ are the photoabsorption cross-section and photoreaction efficiency as a function of wavelength \citep[e.g.,][]{Huebner2015}. The prefactor $a_{\theta}$ is equal to unity if consider the heating at the substellar point, or equal to $ \mathrm{cos}(\SI{66}{\degree})$ to  account for global averaging of the XUV flux \cite{Johnstone2019}. The polytropic model heating per unit mass at the sonic point is given by
\begin{equation}
    q_{sc,\gamma} = \left( \frac{\gamma_a}{\gamma_a - 1} + \frac{\gamma}{1 - \gamma} \right) \frac{4+\sqrt{2} \sqrt{5-3 \gamma}}{(1+\gamma) \cdot (1-\gamma)^{-1}} \frac{RT_{sc}u_{sc}}{r_{sc}}, \label{eqn: heat_sonic}
\end{equation}
calculated from substituting the sonic density temperature gradient, found by chain rule from Equation (\ref{eqn: scale-height}), into Equation (\ref{eqn: poly-heat}). Again, the isothermal limit of $q_{sc}= 3 u_{sc}^3 / r_{sc}$ can be recovered. The XUV flux found by matching Equation (\ref{eqn: heat_thin}) with (\ref{eqn: heat_sonic}) will not necessarily agree with the onset flux (Eqn \ref{eqn: fluxon}). There is broad agreement in the cases we explore; however, the atmospheric structure can be useful regardless. We will now discuss the physical inputs for Equation (\ref{eqn: fluxon}).

\begin{deluxetable*}{ccccc}
\tablecaption{Nitrogen Photoabsorption Reactions \label{tab: photo-abs}}
\tablehead{
\colhead{Photoreaction} & \colhead{$\Delta E$ (eV)} & \colhead{$\lambda_{\nu}$ (nm)} & \colhead{Excess energy (eV)} & \colhead{$\eta_{pr}$}}
\startdata
N$_{2}$ + $\gamma \to$ N$_{2}^+$ + $e^{-}$ & 15.7 & 79.2 & 21.4 & 0.58\\
N$_{2}$ + $\gamma \to$ N + N & 12.5 & 99.2&  3.38 & 0.21\\ 
N$_{2}$ + $\gamma \to$ N$^+$ + N + $e^{-}$ & 24.4 & 25.0 & 32.4 & 0.57 \\ \hline
N + $\gamma \to$ N$^{+}$ + $e^{-}$ & 14.6 & 85.10 & 19.0 & 0.57 \\ \hline 
N$^{+}$ + $\gamma \to$ N$^{2+}$ +$e^{-}$ & 29.2 &  42.5 & 17.3 & 0.37 \\  \hline \hline 
\enddata
\tablecomments{$\Delta E$ \& $\lambda_{\nu}$ are the threshold energy and corresponding wavelength of incoming EUV photons for the given pathway. The excess energy is the average photon energy above the dissociation or ionization threshold and has been calculated in the Modern Solar Spectrum in a period of high activity with data taken from PHIDRATES database \cite{Huebner2015}. The photoreaction efficiency $\eta_{pr}$ is calculated as excess energy divided by the sum of excess energy and threshold energy.}
\end{deluxetable*}

\subsection{Photoevaporation of Nitrogen Thermospheres by Active Stars}\label{subsec: nitrogen}
Differential rotation in a star's convective envelope generates a magnetic field that confines and heats coronal plasma to millions of kelvin, resulting in XUV emission from metal lines and the free-free continuum. Stars are initially saturated in X-ray emission relative to bolometric luminosity, later losing angular momentum through magnetic breaking of the stellar wind until converging onto a main-sequence activity track for a given mass \citep{Vilhu1984}. Depending on the initial rotation rates of F, G, and K stars, their saturated phase lasts around $10-300$ million years, compared to billions of years for the lower mass M dwarfs due to their persistently long convective turnover timescale \citep{JohnstoneFGKM}. 

The evolution of the Sun's XUV flux at 1 AU \citep{Catling&Kasting,Ribas_2005} has been estimated as
\begin{equation}
    F_{\mathrm{xuv}}(\tau_{\text{age}}) \approx 23.3 \times 10^{-3} \tau_{\text{age}}{}^{-1.23} \mathrm{~W} \mathrm{~m}^{-2},
\end{equation}
where $\tau_{\text{age}}$ is in GYrs, $F_{\scriptscriptstyle \mathrm{XUV}, \earth}\approx$ \SI{4e-3}{\watt \per \m \squared}. The XUV flux in the Sun's early pre-main-sequence is then $F_{\scriptscriptstyle \mathrm{XUV}}(\earth, 0.1 \mathrm{~GYr})\approx$ \SI{0.4}{\watt \per \m \squared} $\approx 100F_{\mathrm{xuv}, \earth}$, but could have been several times higher if initial rotation was rapid \citep{JohnstoneFGKM}. Variability in the solar cycle leads to a factor of 2 difference between the EUV of the quiet and active Sun \citep{Huebner2015}. Turning to a $0.09M_{\sun}$ M dwarf, the XUV flux at TRAPPIST-1 b (0.011 AU) during its early pre-main-sequence may have been $\gtrsim 10^4 F_{\scriptscriptstyle \mathrm{XUV}, \earth}$ and today it is $ \sim 10^3 F_{\scriptscriptstyle \mathrm{XUV}, \earth}$ \citep{Fleming_2020}. While for the M dwarf super-Earths LHS 1140 c (0.027 AU) and LHS 1140 b (0.09 AU) the XUV fluxes today are approximately $100 \times$ and $ 10 \times F_{\scriptscriptstyle \mathrm{XUV}, \earth}$, with b currently in the habitable zone \citep{Spinelli_2023, Cadieux_2024}.  

At the base of the thermosphere, photodissociation of dinitrogen produces atomic nitrogen that is advected upwards, while N$_2$ is replenished from the lower-atmosphere reservoir. For an energy-limited onset  we expect that photoionization of atomic nitrogen is balanced by the advection of ions downstream along a gradient of increasing ionization fraction. Ion-electron recombination plays a minor role in these conditions (see Sec \ref{sec: nlte}). The flow accelerates sharply, with density dropping to keep mass flux roughly uniform. We expect a relatively sharp N$_2$/N front followed by an extended N/ N$^+$ front up to the sonic point where, with onset XUV fluxes, the ionization fraction should remain minor, as is found in \cite{Johnstone2019}.

The reaction pathways for XUV absorption are summarized in Table \ref{tab: photo-abs}, with data from \cite{Huebner2015}. Photodissociation to neutral atoms does not efficiently heat the thermosphere because absorption is constrained by the potential energy of bound states of dinitrogen \citep[e.g.,][]{Leiden}. To find the global efficiency of heating by photoreactions, when ion-electron recombination and dissociative recombination of nitrogen are low, one can make the approximation that every two atoms lost must have come from a photodissociation event and every ion lost from a photoionization event. Thus, weighting the average excess energies, the global photoreaction heating efficiency is $\eta_{pr}(f_{+}^{sc}=0.1) \approx 0.21 + 2\times 0.1\times 0.57 \approx 1/3 $, where $f_{+}$ is the ionization fraction. For comparison, this photoreaction efficiency is approximately equal to the EUV heating efficiency found by \cite{France_2020} for typical Earth-Sun conditions. We note that the efficiency of the \textit{local} heating at the sonic point is determined by excess photoionization energy only, yielding $\eta_{pr, sc} \approx 0.6$.

Oxygen has a lower dissociation energy than nitrogen, with absorption extending in the UV to $\sim $\SI{200}{\nm}. The model of \cite{Johnstone2019} includes UV in their XUV spectrum, tripling the energy available in their $\approx 60 F_{\scriptscriptstyle \mathrm{XUV}, \earth}^*$. Focusing attention on a pure nitrogen atmosphere allows us to safely neglect detailed photochemistry and means the XUV spectrum mostly overlaps with that driving hydrogen escape. Apart from order unity variation in efficiency, the polytropic heating is agnostic to what composes the XUV spectrum, so fluxes quoted in Wm$^{-2}$ are unaffected. However, the dynamics of oxygen line cooling compared to nitrogen is more heterogenous so we will make direct comparison in Section \ref{sec: nlte}.

The neutral collision cross-section given by $\sigma_{\scriptscriptstyle C}(\mathrm{N-N}, 10^4 \ \mathrm{K})=$ \SI{3.9e-20}{\m\squared} \citep{Kislyakova}. Charge-exchange is effective enough that even for a weakly ionized plasma, it could more than double the total cross-section: $\sigma_{ \scriptscriptstyle C}(\mathrm{N-N+}, 10^4 \mathrm{K})=$ \SI{5e-19}{\m\squared} \citep{Laricchiuta2009}. The electron-neutral cross section can be found somewhere inbetween $\sigma_{\scriptscriptstyle C}(\mathrm{N-e^{-}}, 10^4 \ \mathrm{K})=$ \SI{2e-19}{\m\squared}. For a species mixture, an effective collision cross-section against the bulk density can be considered: 
\begin{equation}
    \sigma_{ \scriptscriptstyle \mathrm{C}} = \sum_{k} \frac{n_{k} \sigma_{ k, \scriptscriptstyle{\mathrm{C}} }}{n}.
\end{equation}
Considering an ionization fraction $\sim 0.1$ we find $\sigma_{\scriptscriptstyle C}(\mathrm{N-all}, 10^4 \ \mathrm{K})=$ \SI{1.1e-19}{\m\squared} $\sim 3\times\sigma_{\scriptscriptstyle C}^{\mathscr{N}}$.
\subsubsection{Young Sun Earth} \label{subsec: YSE}
\begin{figure}[htbp!]
  \centering
  {\phantomsubcaption\label{fig:4A}}
  {\phantomsubcaption\label{fig:4B}}
  {\phantomsubcaption\label{fig:4C}}
  \tikz\node[inner sep=2pt,label={[anchor=north west]north west:\subref{fig:4A}}]{\includegraphics[width=0.9\linewidth]{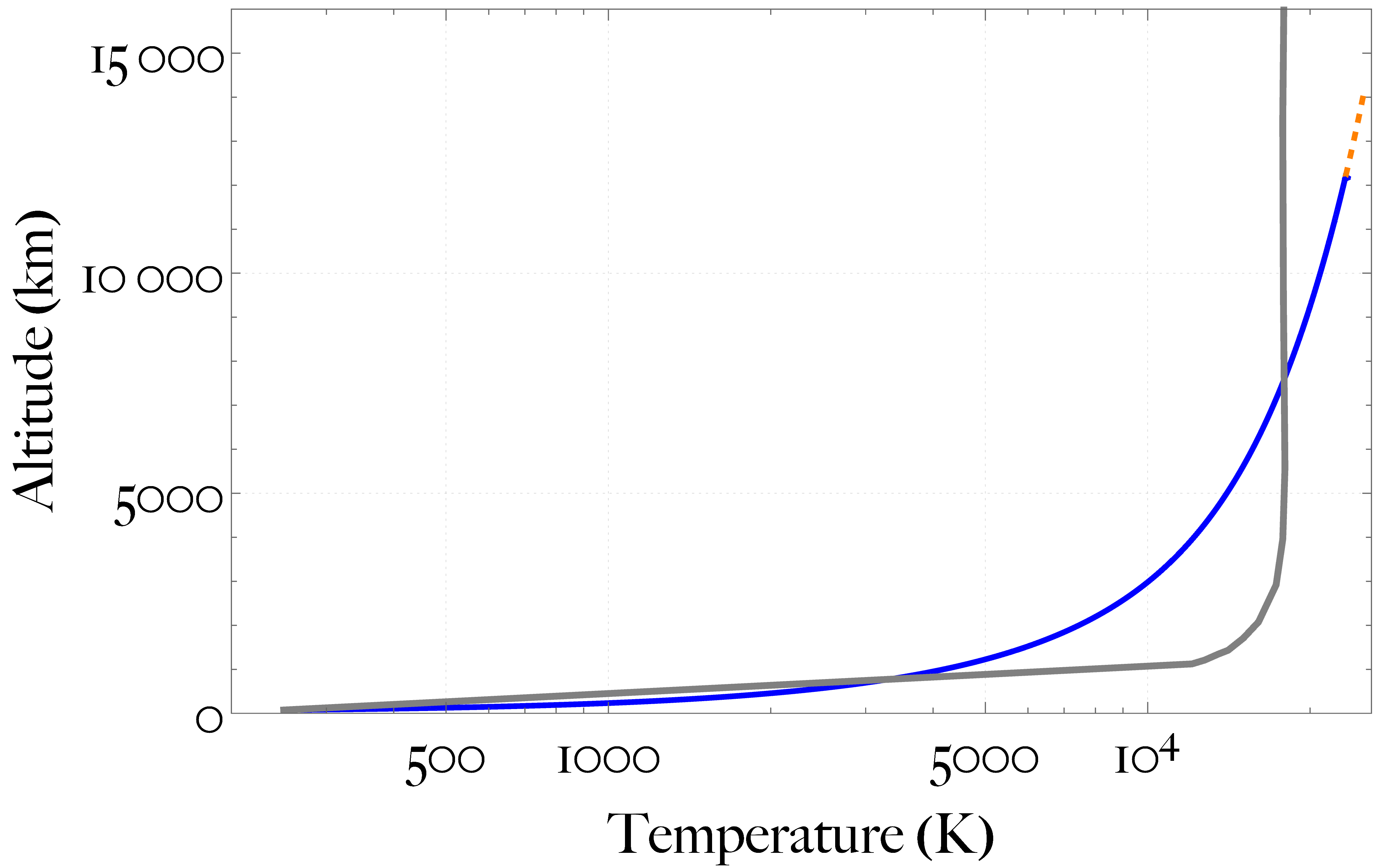}};
  \tikz\node[inner sep=2pt,label={[anchor=north west]north west:\subref{fig:4B}}]{\includegraphics[width=0.9\linewidth]{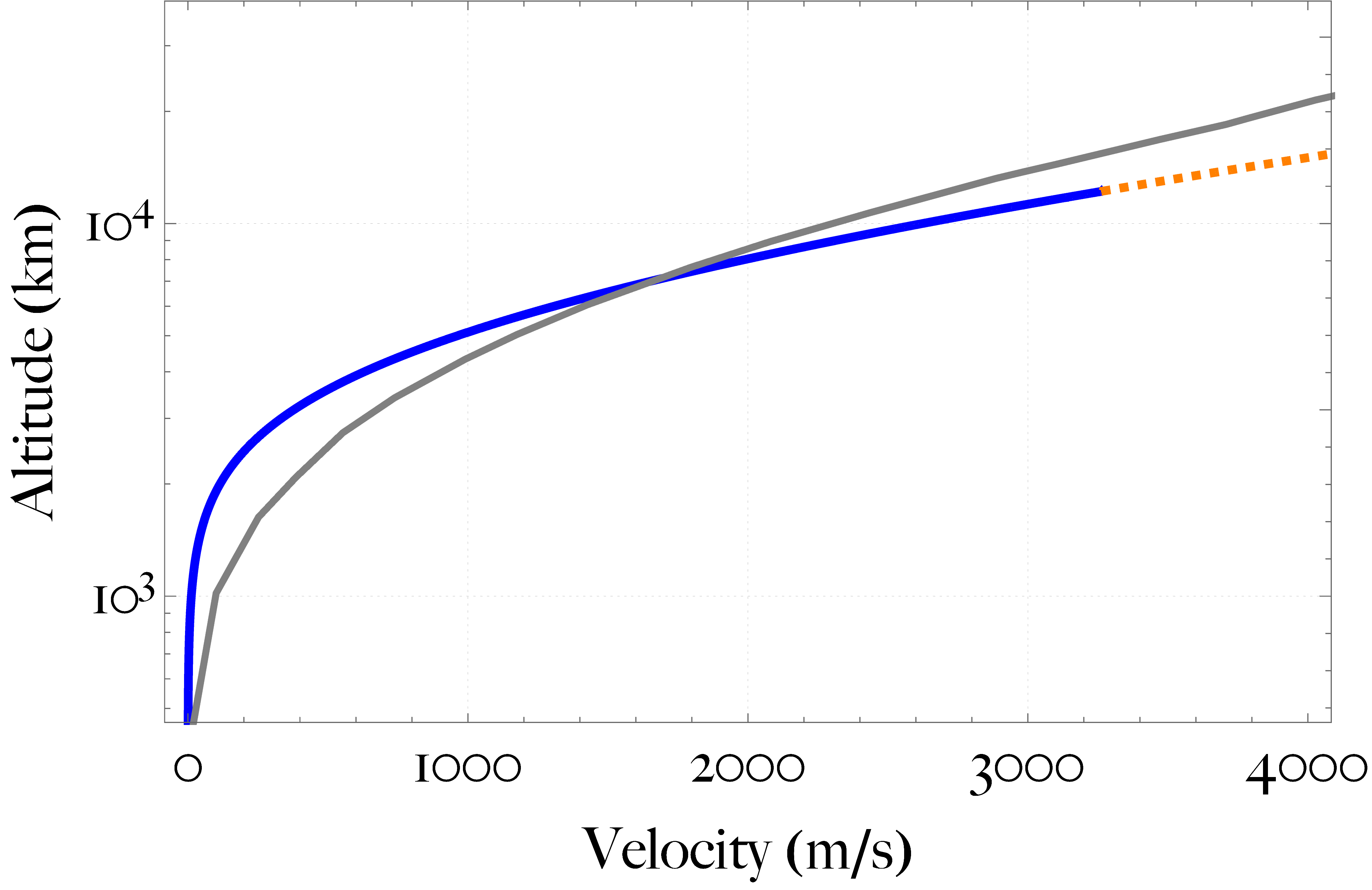}};
  \tikz\node[inner sep=2pt,label={[anchor=north west]north west:\subref{fig:4C}}]{\includegraphics[width=0.9\linewidth]{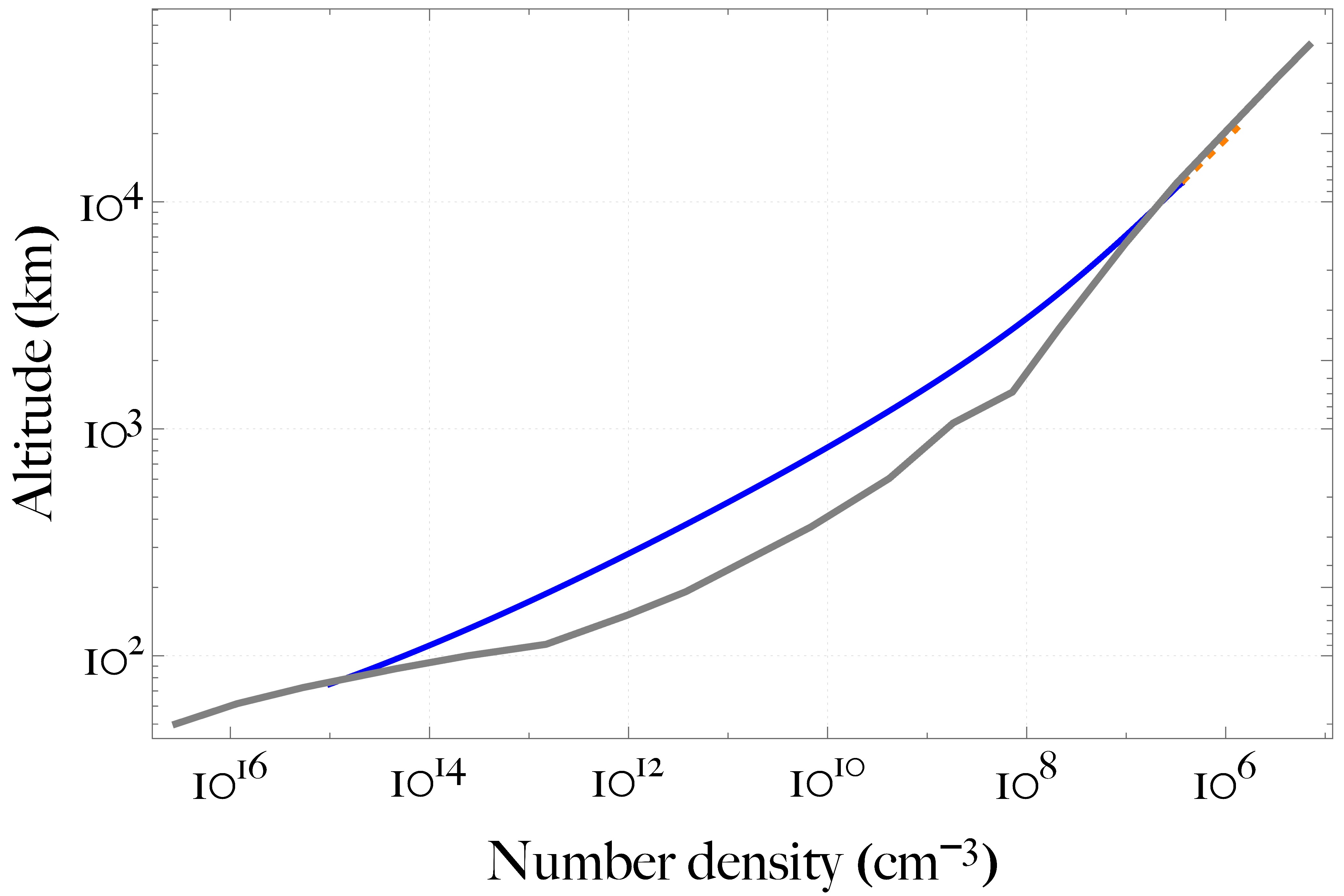}}; \\
  \caption{The polytropic onset solution $(\lambda_0, \gamma_{\scriptscriptstyle \mathrm{on}}) = (417,0.77)$  for the temperature, velocity and number density profiles (blue) for an Earth-like planet hosting a pure nitrogen thermosphere. The dashed-orange line begins at the sonic level $r_{sc} \approx 2.9 R_{\earth}$. The hydrocode profiles from \cite{Johnstone2019} are presented for comparison in grey. The XUV flux required to drive the transonic hydrodynamic escape is estimated $ F_{\scriptscriptstyle \mathrm{XUV}}^{\mathrm{on}}(\mathrm{YSE})\approx$ \SI{0.84}{\watt\per\m\squared}, or equivalently, $\gtrsim 200 F_{\scriptscriptstyle \mathrm{XUV}, \earth}$.} \label{fig: YSE-structure}
\end{figure}

This section presents a model of the Knudsen onset transonic outflow of Earth's atmosphere prescribed to be pure nitrogen, which may have been energetically feasible in the raised XUV output of the young Sun. Though the Early Earth is thought to have had a CO$_2$-rich atmosphere \citep[e.g.,][]{Lammer2018}, nitrogen's role in habitability makes this an interesting exo-Earth scenario. Somewhat coincidentally, the escape rate of the \cite{Johnstone2019} hydrodynamic simulations of young Sun Earth turns out to be close to the neutral onset that we here derive from the polytropic framework, offering the opportunity for intercomparison of  atmospheric profiles. Though we note that the collisional onset within the hydrocode model could be different, depending on collision cross-section, the scale height and definition of the sound speed. 

The lower boundary is set to a thermobase-like altitude of \SI{75}{\km} with an inventory of $10^{15}\mathrm{cm^{-3}}$ nitrogen atoms and a temperature of $250$\,K, yielding $\lambda_0 = 416$ noting $M_{\earth} =$\SI{5.97e24}{\kg} and $R_{\earth}=$\SI{6.378e6}{\m} (equatorial). Calculated with Equation (\ref{eqn: kin_poly}), the neutral onset  $\gamma_{\mathrm{on}}=0.77$ provides enough implicit heating for the scale height to develop such that the outflow remains collisional out to the sonic point. This corresponds to almost $9$ orders of magnitude change in number density from the base to the sonic radius (Fig \ref{fig: YSE-structure}). The sonic radius is calculated as $2.9\times$ the planetary radius (Eqn \ref{eqn: itr}), or at an altitude of approximately \SI{12200}{\km}. The temperature monotonically increases to \SI{23300}{\kelvin} at the sonic radius, yielding a sonic-to-base temperature ratio of $93\times$ (Eqn \ref{eqn: Tsonic}). The velocity is inappreciable until a couple of thousand kilometers in altitude before rising steeply to the sonic speed at \SI{3.3}{\km \per \second} --- the total change in Mach number is seven orders of magnitude. Past the sonic radius, a dashed line shows a continuation of the polytropic solution for each profile in Figure \ref{fig: YSE-structure}.  However, the behaviour and forcing downstream of the sonic point do not affect the mass loss rate or the atmospheric structure upstream.

The polytropic model for young Sun Earth onset (Eqns \ref{eqn: mass loss} \& \ref{eqn: kin_poly}) yields a mass loss rate of \SI{1e9}{\gram \per \second} or \SI{4e31}{\text{atoms} \per \second}. Based on this rate, six bars of an $\mathrm{N}_{2}$ atmosphere could be lost in a million years. As noted by \cite{Johnstone2019}, the discussion of fluxes is always up to an order unity factor as studies use different averaging conventions and a range of spectral models. The flux-receiving cross-section has a larger area than the projected surface of the planet ($\pi R_{pl}^2$) - larger due to heating being significant out to a few planetary radii. We approximate the geometric factor $R_{abs}= \sqrt{2} R_{pl}$ (Eqn \ref{eqn: E-L}) following \cite{Catling&Kasting}. We also account for the thermomechanical efficiency of the hydrodynamic expansion $\eta_{0.77} \approx 0.76$ and the efficiency of heating via photoelectrons $\eta_{pr} \approx 1/3$. Altogether, we find the flux incident on the sub-stellar column needed to trigger the onset of transonic hydrodynamic escape (Eqn \ref{eqn: fluxon}) to be $ F_{\scriptscriptstyle \mathrm{XUV}}^{\mathrm{on}}(\mathrm{YSE})\approx$ \SI{0.84}{\watt\per\m\squared}, equivalent to $\gtrsim 200 F_{\scriptscriptstyle \mathrm{XUV}, \earth}$. However, a subsonic onset of bulk outflow may be possible for fluxes an order of magnitude lower $\sim 20F_{\mathrm{xuv}, \earth}$.  Accounting for the uncertainty in the initial rotation rate of the Sun \citep{JohnstoneFGKM} and neglecting non-LTE cooling for now, the transonic escape of a thick secondary atmosphere during the saturated phase is plausible. 

The $\gamma_{\mathrm{on}}=0.77$  mass loss rate happens to be $1/2 \times$ the rate in \cite{Johnstone2019}, so broadly similar profiles might be expected. Figure \ref{fig: YSE-structure} displays the curves calculated with \texttt{KOMPOT} from \cite{Johnstone2019} for comparison. A limitation of the polytropic models is the neglecting of the photodissocation front, instead considering an inventory of nitrogen atoms. Consequently, the mean molecular weight in the first few scale heights can be underestimated to as low as half of its true value. However, the steeper-then-flatter pattern converges to a similar density where the polytropic atmosphere goes sonic to within a factor of two. 

The temperature profiles are similar, but the comparison highlights that a single polytropic index cannot capture the differential steepness of the middle and upper atmospheres, as the photodissociation heating peak is neglected. Furthermore, although terminal atmospheric outflows are commonly isothermal or exhibit adiabatic cooling because advection balances or exceeds local heating, the polytropic solutions reported here all maintain sustained temperature inversions. As a result, sonic temperatures are elevated compared to the escape temperature $T_{sc}(r_{sc})= T_{esc}(r_{sc})/\gamma$; see Eqns (\ref{eqn: Tesc}) \& (\ref{eqn: Tsonic}). The $\gamma_{\mathrm{on}}=0.77$ inversion is steep, so the sonic temperature of \SI{23300}{\kelvin} is likely an overestimate relative to the roughly isothermal terminal temperature of $\sim$\SI{18000}{\kelvin} reported by \cite{Johnstone2019}, However, it remains physically self-consistent with a strongly heated upper atmosphere. When determining the saddle point of the momentum equation, increasing  heating local to the sonic point requires an increase in temperature to keep the left-hand side of Equation~\ref{eqn: transonic rule} equal to zero. \cite{lamers1999} explained this perhaps counter-intuitive result: local heating produces an inward-directed force that must be balanced.

Altogether, the polytropic approximation naturally produces, without fitting, profiles that bear key similarities to a detailed hydrocode solution \citep{Johnstone2019} at a similar XUV flux. Had we attempted an isothermal Knudsen onset (Eqn \ref{eqn: kin_iso}), we would have faced the $(\lambda_0, n_0)$ degeneracy, or the problem of how to choose the sonic radius. For the thermobase density of $10^{15}\mathrm{cm^{-3}}$, the isothermal Knudsen onset (Eqn \ref{eqn: kin_iso}) would find an almost unrecognizable solution: $ r_{sc} = \frac{\lambda_0}{2} R_{pl} \approx 11 R_{pl} $, $T \approx 4100$\, K --- leading to a mass loss rate about $5\times$ higher than the polytropic case. Though reducing the base density of the Parker wind would allow a hotter and less extended outflow to maintain $\mathrm{Kn}_{sc}=1$,  one must then use ad hoc fitting to produce a realistic profile.

\subsubsection{Young Sun Mars}
\label{subsec: Mars-onset}
Following the last section, we now present a model of evaporation of a nitrogen from Mars in the raised XUV output of the young Sun. Early Mars may have had more than $0.1$ bars of nitrogen \citep{Hu2022}. However, our primary motivation is to gain insight into exo-Mars planets and the landscape of high molecular weight escape. Follow-up work of coupling the polytropic model to a chemical kinetics code with a CO$_2$ photochemistry network is in progress (Blumenthal et al., in prep).

\begin{figure}[htbp]
\includegraphics[scale=0.38]{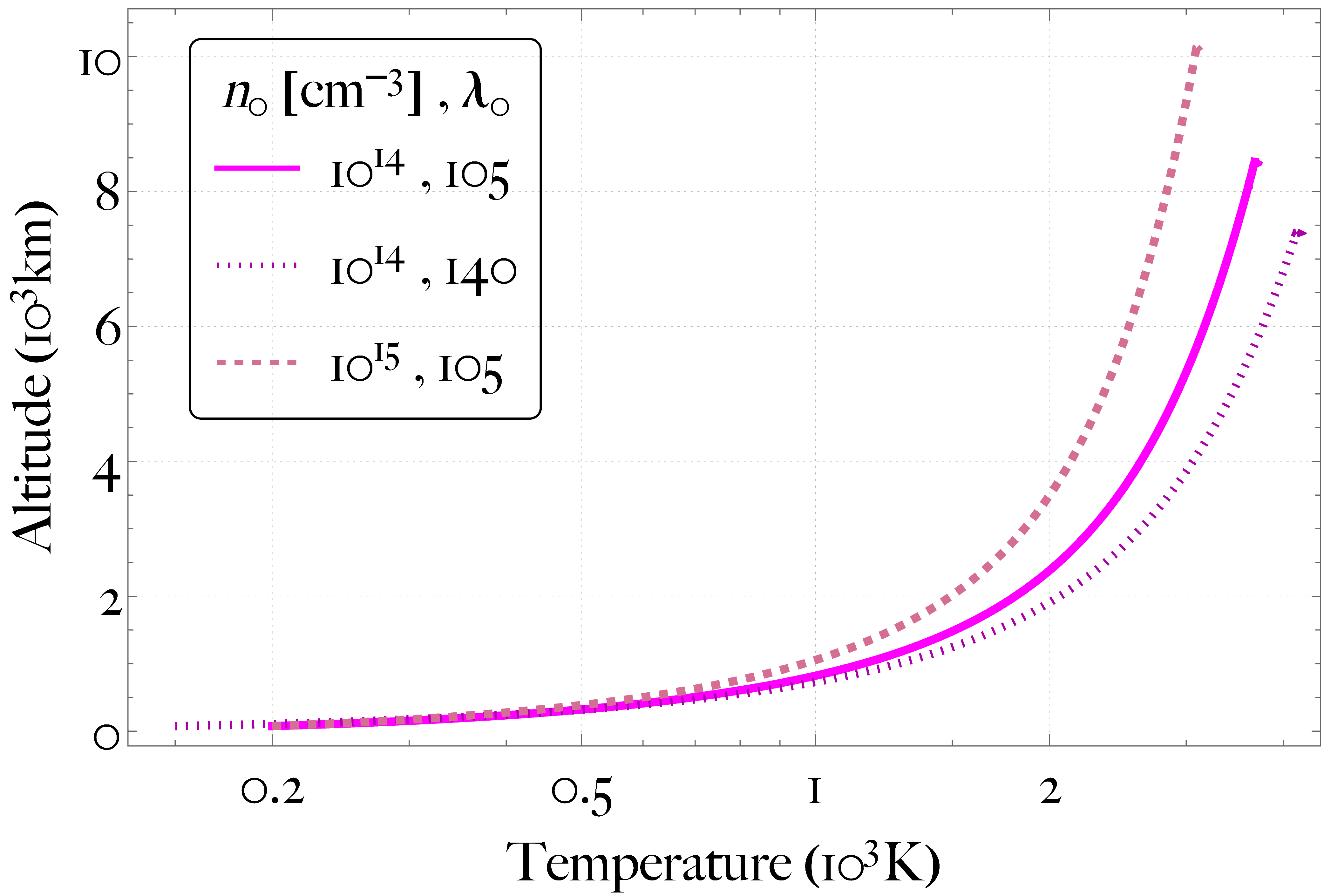}
\caption{Temperature profiles for hypothetical nitrogen escape from young Sun Mars $(T_0, n_0, \lambda_0, \gamma_{\scriptscriptstyle \mathrm{on}})=$ (\SI{200}{\kelvin},  \SI{e14}{\per\cm\cubed}, 105, 0.826) (solid), (\SI{150}{\kelvin}, \SI{e14}{\per\cm\cubed}, 140, 0.8) (dashed),   (\SI{200}{\kelvin}, \SI{e15}{\per\cm\cubed}, 105, 0.86) (dotted). } \label{fig: marsT} 
\end{figure}

For a base temperature \SI{200}{\kelvin}, noting $M =$ \SI{6.39e23}{\kg} and $R=$ \SI{3.39e6}{\m} (equatorial), the escape parameter for Mars is $\lambda_0 =$ 105. For a thermobase inventory of \SI{e14}{\per\cm\cubed} nitrogen atoms, Equation (\ref{eqn: kin_poly}) yields $\gamma_{\mathrm{on}}=0.826$. Reducing the base temperature from \SI{200}{K} to \SI{150}{K} increases $\lambda_0 =$ 140 and a steeper inversion of $\gamma_{ \mathrm{on}}=0.80$ is required to remain collisional out to the sonic point. On the other hand, increasing the base density to \SI{e15}{\per\cm\cubed} allows a less steep inversion of $\gamma_{\mathrm{on}}=0.86$. For $\gamma_{\mathrm{on}}=0.826$, the sonic radius is $3.5\times$ the planetary radius or at an altitude of approximately \SI{8500}{\km} (Fig \ref{fig: marsT}). The temperature monotonically increases to \SI{3600}{\kelvin} at the sonic radius, yielding a sonic-to-base temperature ratio of $18\times$. 

The indicative young Sun Mars transonic onset (Eqn \ref{eqn: fluxon}) yields a mass loss rate of \SI{3e8}{\gram \per \second} or \SI{1.3e31}{} atoms per second or 2 bars of N$_2$ per million years. Using the same efficiency factors as from the Earth model, the flux incident on the sub-stellar column $ F_{\scriptscriptstyle \mathrm{XUV}}^{\mathrm{on}}(\mathrm{YSE})\approx$ \SI{0.2}{\watt\per\m\squared}, equivalent to $\gtrsim 50 F_{\scriptscriptstyle \mathrm{XUV}, \earth}$ or $ 100 \times$ Modern Day Mars XUV. The sonic temperatures in the $\gamma_{\mathrm{on}}=0.80$ and $\gamma_{\mathrm{on}}=0.86$ cases are \SI{3100}{\kelvin} and \SI{4200}{\kelvin}, respectively. The resultant difference in mass loss is less than 10 \% due to the square-root dependence of the sonic velocity on temperature, showing a relative insensitivity to the lower boundary conditions. While for the isothermal Knudsen onset, if we halve the (base) temperature, the sonic radius doubles, the mass loss rate increases by $\sqrt{2}$, and the ratio of base-to-sonic density is increased orders of magnitude by a factor of e$^{2r_{sc}/R_{pl}}$.

\cite{TianMars} found hydrostatic instability and rapid escape of a CO$_2$ atmosphere on Early Mars for $\sim 20 \times$ Modern Day Mars XUV. However, this is not inconsistent with the higher XUV required for transonic onset reported here because subsonic outflow could be driven at those lower fluxes. In conclusion, the hypothetical transonic loss of a nitrogen atmosphere from Mars in the saturated phase of the young Sun would have been possible. Furthermore, the atmosphere of an Exo-Mars in the Habitable Zone of an M dwarf is not expected to survive on geological timescales. 

\begin{figure*}[htpb!]
\includegraphics[scale=0.75]{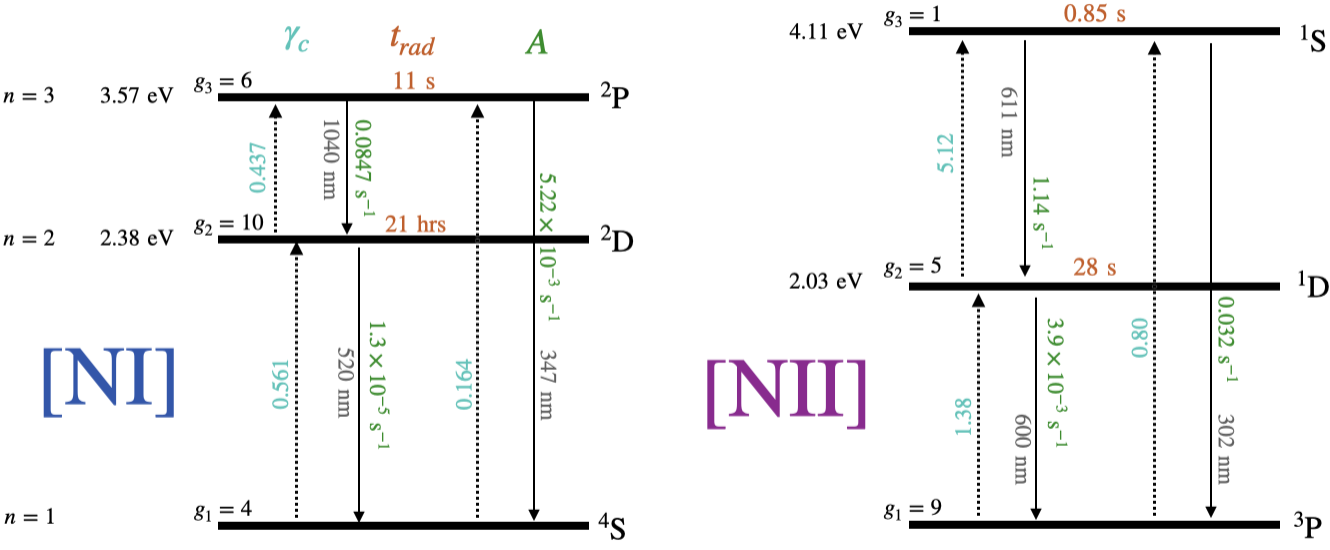}
\caption{Grotrian level diagram for the first three levels of neutral and singly ionized nitrogen. The coefficients for the collisional excitation $\gamma_{\scriptscriptstyle C}$, the radiative lifetime of the level $t_{rad}$ and the transition Einstein-$A$ from \cite{Nakayama_2022} are displayed.}  \label{fig: grotrian}
\end{figure*}

\subsection{\texorpdfstring{Comparison with \cite{Johnson2013, Johnson2013b}}{Comparison with Johnson 2013 and Johnson 2013b}}
\label{subsec: Johnson}
The analytic transonic criterion of \cite{Johnson2013} was employed by \cite{Gronoff} to evaluate the propensity for atmospheric escape across a survey of planets. Notably, the Direct simulation Monte Carlo models from \cite{Johnson2013} suggest that the energy-limited mass loss rate applies well at onset and for a range of subsonic flow at smaller XUV fluxes. However, \citep{Johnson2013} explore a limited set of assumptions and regimes, so determining the general extent to which the energy limit remains valid before transitioning non-linearly to a roughly Jeans-type escape remains a key unresolved question in the field. On comparing Equation (\ref{eqn: fluxon}) with "Equation (10)” from \cite{Johnson2013}, we see that each calculation estimates the total net heating required to drive transonic outflow at an order-unity Knudsen number at the sonic point. For young Sun Earth, \cite{Johnson2013, Johnson2013b} identify the transonic onset at $\gtrsim 100 F_{\scriptscriptstyle \mathrm{XUV}, \earth}$, consistent with our results. The polytropic model presented here further enables analytic determination of the sonic scale height and velocity, as well as evaluation of the thermomechanical and line cooling efficiency from the atmospheric structure. We also incorporate a critical estimate for the photoreaction efficiency factor of a nitrogen atmosphere (see Table \ref{tab: photo-abs}). In subsequent sections, we refine the transonic criterion by analyzing the breakdown of neutral Knudsen-onset assumptions and mapping the landscape of high molecular weight escape (see Fig \ref{fig: conceit}).

\section{Accounting for Non-LTE Cooling} 
\label{sec: nlte}

The previous section discussed the conditions required for the transonic escape of secondary atmospheres for Mars-to-Earth-sized planets while neglecting the effects of forbidden line cooling. We will now test the primacy of atomic line cooling in the evolution of secondary atmospheres found by \cite{Nakayama_2022}, but for the hydrodynamic rather than hydrostatic regime. We introduce the simplifying approximation of a three-level model of electron-impact excitation of forbidden lines that cool directly to space. The non-LTE radiative transfer of molecular lines, shown to be important in escaping steam atmospheres \citep{Munoz2024}, is not discussed here. 
\subsection{Three-level Model for Electron Excitation of Atomic Lines} \label{subsec: LC}

The polytropic framework admits atomic and molecular cooling estimates diagnosed from the atmospheric structure. For a polytropic solution to represent a consistent, steady state, the line cooling rate per unit mass should be lower than the polytropic heating throughout the extent of the atmosphere. The total line cooling luminosity informs the efficiency factor for the energy limit. Collisions between atoms, molecules, electrons, and ions excite vibrational, rotational modes, and electronic states, which can be relaxed by further collisions or by spontaneous emission of light. Electronic states are available to ambient temperatures of \qtyrange{0.1}{1}{\electronvolt}, equivalent to \qtyrange{e3}{e4}{\kelvin}, or greater. In Local Thermodynamic Equilibrium (LTE), the photons are honorary gas particles, meaning that the local emission per unit mass depends only on a single temperature field and is independent of local densities. Maintained by collisions, the well-known Boltzmann distribution gives the population ratio between levels $j$ and $i$: $\frac{n_j}{n_i} = \frac{g_j}{g_i}\exp{-E_{ji}/k_{B} T}$, where $E_{ji}$ is the excitation energy, $g_{i}, g_{j}$ are the level density of states and $k_{B}$ is the Boltzmann constant \citep[e.g.,][]{feynman}. The resulting volumetric cooling rate from some level $j$ to a lower level $i$ via spontaneous emission is given by $Q_{ji}= \frac{n_i g_j}{g_i}\exp{-E_{ji}/k_{B} T} A_{ji} E_{ji}$. 

Atomic excitation is dominated by electron impacts because electron thermal velocities are $\sim\sqrt{m_i/m_e}$ ($\approx 160$ for N) times larger than ion velocities, fast enough to produce sudden (non-adiabatic) excitation of bound electrons \citep{Hertel2015}. In contrast, heavy-particle collisions occur on timescales long compared to the intrinsic electronic timescale ($\sim h/ E_{ji}$), leading to adiabatic motion and an exponential suppression of electronic excitation in N—N or N$^+$—N collisions. Heavy-particle de-excitation, however, is exothermic and can proceed through adiabatic evolution of a transient quasi-molecule. Although such heavy-particle quenching of excited states may reduce the efficiency of atomic line cooling under certain conditions, calculation of the effect is beyond the scope of the present study.

The other density regime is the coronal limit, where collisional excitations are in dynamic balance with spontaneous emission - a simple non-LTE limit. The volumetric cooling rate then becomes independent of $A_{ul}$ in the two-level atom approximation: every electron excited to the upper level spontaneously decays, so that
\begin{subequations}
\begin{eqnarray}
Q_{ul}^{(2)} &=& n_{l} C_{lu} E_{ul},\label{eqn: coronal} \\ 
C_{l u} &=& \gamma_{ \scriptscriptstyle \mathrm{C}} n_{e} \frac{8.629 \times 10^{-6}}{g_l \sqrt{T}} \exp \left(-\frac{ E_{l u}}{k_{B} T}\right), \label{eqn: 2lvl} 
\end{eqnarray}
\end{subequations}
where $C_{lu}$ is  collisional excitation coefficient from the lower $l$ to upper level $u$ and $\gamma_{\scriptscriptstyle \mathrm{C}}$ is the effective collision strength; the formula takes number densities in \SI{}{\cm \cubed} and temperatures in \SI{}{\kelvin} \citep{Nakayama_2022}. An important note is that the two-level coronal cooling scales as the density squared $\propto n_{l} n_{e}$. The two-level critical density $n_{cr}^{(2)}$ separating the LTE and coronal regimes is given by 
\begin{equation}
    n_{cr}^{(2)} = \frac{A_{ul}}{k_{ul}}, 
\end{equation}
where the collisional deexcitation coefficient is
\begin{equation}
C_{ul}= n_e k_{ul} = \gamma_{ \scriptscriptstyle \mathrm{C}} n_{e} \frac{8.629 \times 10^{-6}}{g_u \sqrt{T}},
\end{equation}
and if $n_e \ll n_{cr}^{(2)}$ the plasma will be in the coronal regime. 

In this sub-section, we will explore in detail the states of the neutral nitrogen atom (NI) described by three quantum numbers: principal, orbital angular momentum and intrinsic angular momentum. This accounts for interparticle electrostatics through spin-spin and orbit-orbit coupling but neglects spin-orbit coupling and the associated fine structure.  The first three levels of NI are displayed in Figure \ref{fig: grotrian} adopting term notation, along with those of NII. All data are taken from \cite{Nakayama_2022}, including those of the oxygen lines in OI and OII, which we will sometimes reference for comparison. 

The longest wavelength electric-dipole-allowed transition to ground is a strong line called the resonance line. The NI resonance line is ${}^4 P\to {}^4 S$ with $E_{41} =$ \SI{10.3}{\electronvolt}, $\lambda_{\nu}^\mathscr{O}=$ \SI{120}{\nano\m} and a radiative lifetime of just \SI{3}{\nano\second}. 
However, terrestrial upper atmospheres rank, in terms of astrophysical line sources, as warm $\sim$ \SI{e4}{\kelvin} and tenuous $n_{e} \ll$ \SI{e10}{\per\cm\cubed}, so the resonance line is dramatically underactivated compared to when in hot and dense conditions that make up LTE. Consequently, the low-lying, electric-dipole-forbidden transitions, which are shown in Figure \ref{fig: grotrian}, dominate. At temperatures up to \SI{20000}{\kelvin}, a three-level atom is sufficient to model the collisional-radiative non-equilibrium explicitly. The first two excited states have long radiative lifetimes, but in the three-level coronal limit, the cooling contribution is only weakly dependent on the Einstein-A. 

\begin{figure}[htpb!]
\centering
\includegraphics[scale=0.38]{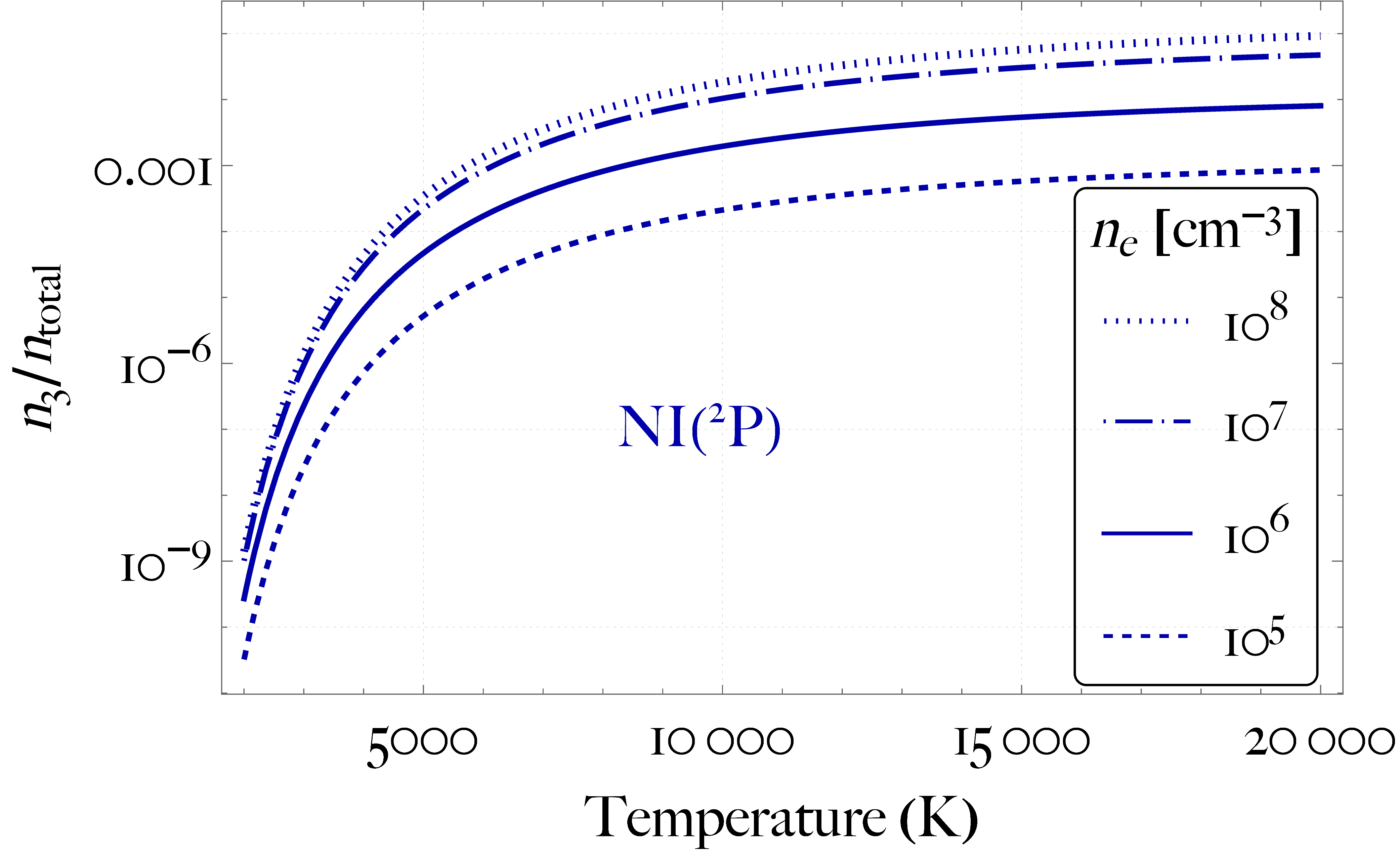}
\caption{The ratio of nitrogen atoms excited two levels above ground $n_{3}$ to the total population $n_{total}$ plotted as a function of temperature and electron density; calculated with a three-level atom approximation (Eqn \ref{eqn: 3lvl}). Note that the cooling luminosity in each line is directly proportional to the number of atoms in each excited level. \label{fig: linecooling}}
\end{figure}

We can approximately model $\left[\mathrm{NI}\right]$ as a three-level atom excited by electron impact, as has been done for fine structure of the NII ground state by \cite{Goldsmith2015}. The populations of these three levels depend on the collisional and radiative Einstein coefficients such that
\begin{align}
-\left(A_{32}+C_{32}+C_{31}\right) n_3+C_{23} n_2+C_{13} n_1 & =0  \notag \\
\left(A_{32}+C_{32}\right) n_3 -\left(A_{21}+C_{21}+C_{23}\right) n_2+C_{12} n_1 & =0  \notag \\ 
C_{31} n_3+\left(A_{21}+C_{21}\right) n_2-\left(C_{12}+C_{13}\right) n_1 & =0  \notag \\ 
n_3+n_2+n_1 & =n_{total}
 \label{eqn: 3lvl}
\end{align}
Reduction of these equations to the population ratios yields a dependence only on the electron density and temperature as displayed in Figure \ref{fig: linecooling}. The three-level critical densities differ from the two-level model. The population of the third level only begins to be independent of density around $n^{cr}_{3} \sim$ \SI{e8}{\per\cm\cubed}, which is about an order of magnitude greater than that found with the inadequate two-level model. We have neglected photochemical production of excited states which should be reasonable for nitrogen atmospheres but might be major in a carbon dioxide atmosphere. 

\begin{figure*}[htbp!]
     \centering
     \begin{subfigure}[b]{0.49\textwidth}
         \centering 
         \caption{Young Sun Mars}     \includegraphics[width=\textwidth]{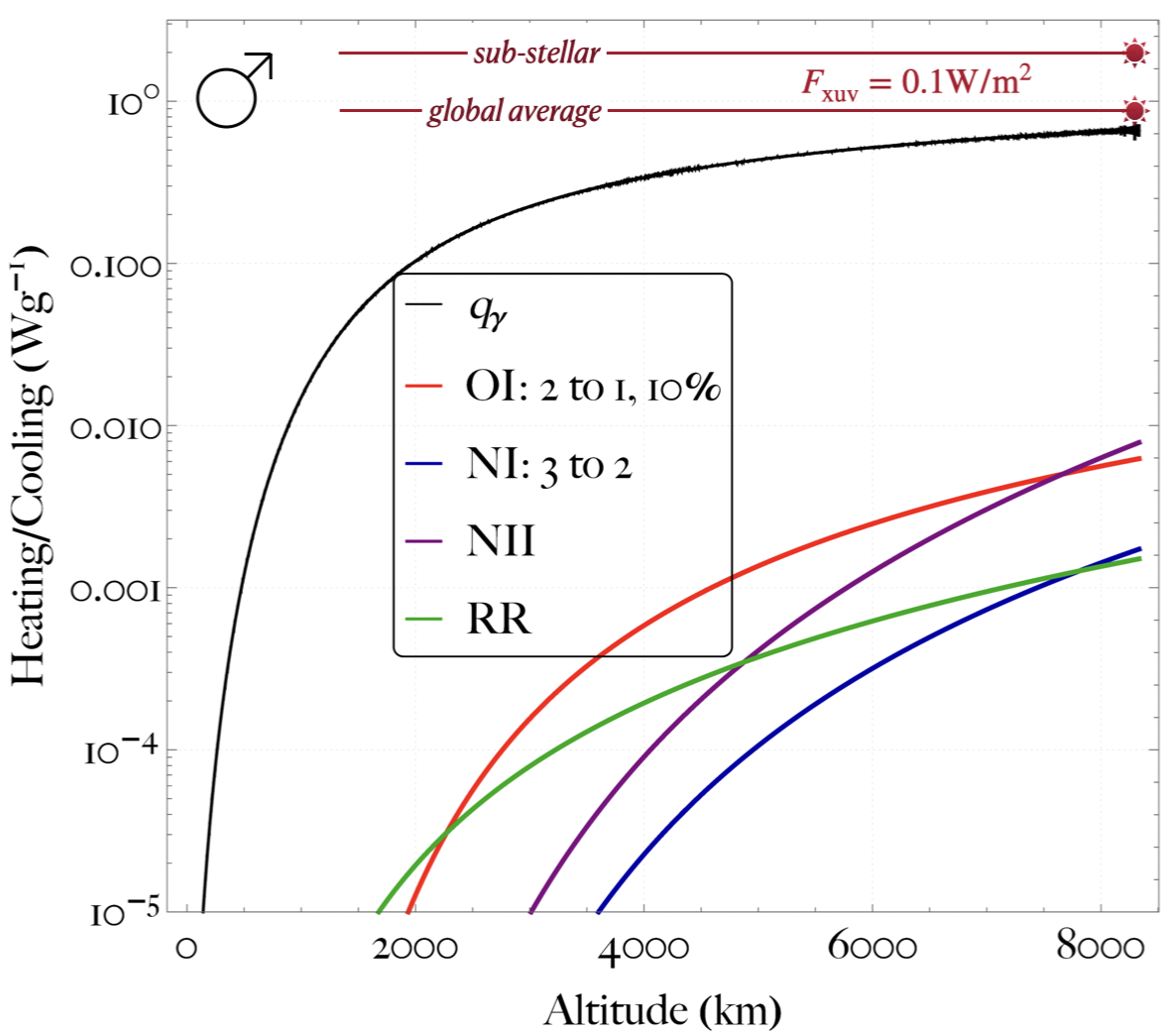} \label{fig: marsHC}
     \end{subfigure}
     \hfill
     \begin{subfigure}[b]{0.49\textwidth}
         \centering 
         \caption{Young Sun Earth}
         \includegraphics[width=\textwidth]{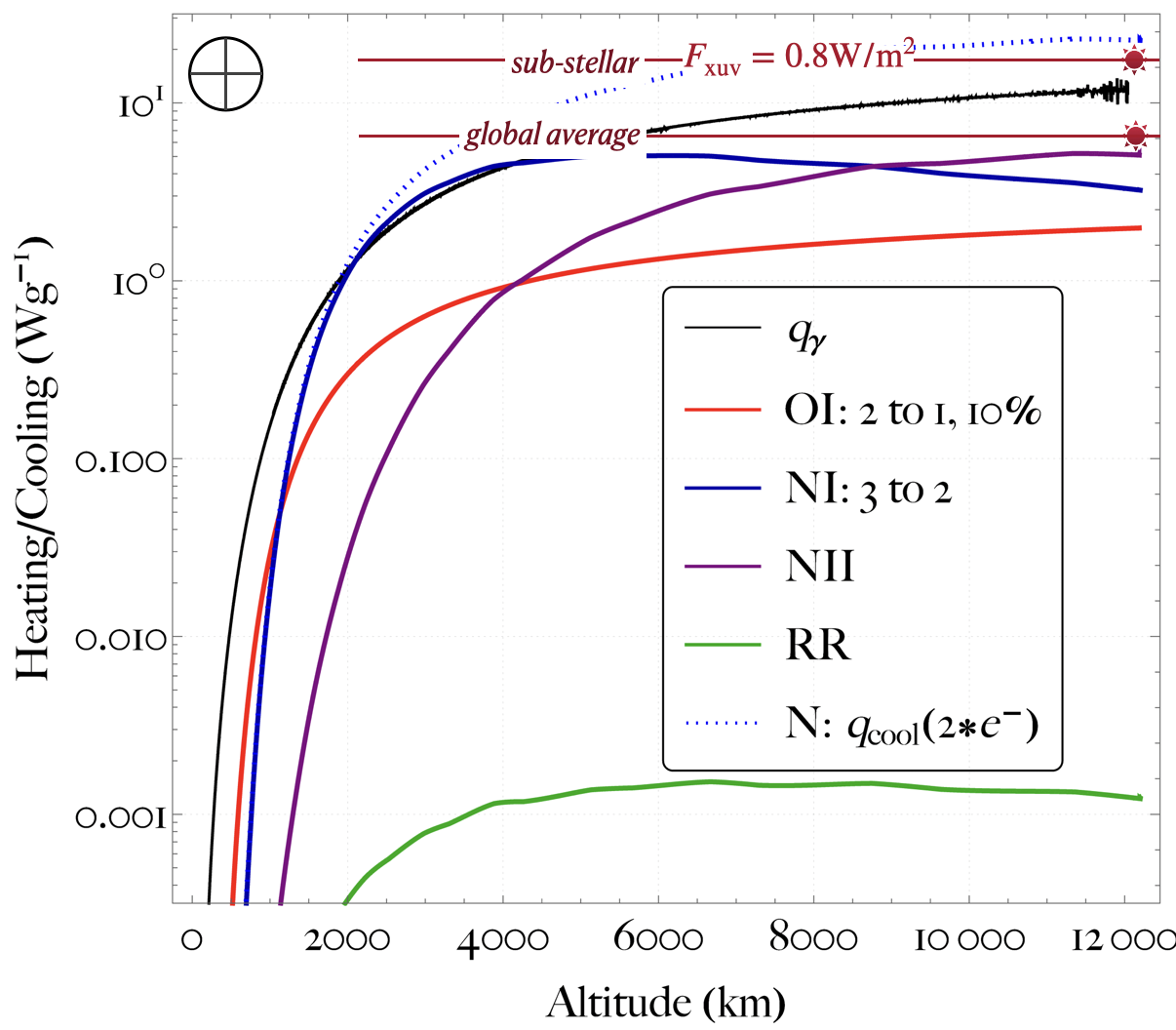}
         \label{fig: YSEHC}
     \end{subfigure}
        \caption{Polytropic heating ($q_{\gamma}$ in black; see Eqn \ref{eqn: poly-heat}) compared to forbidden line cooling (blue, red and purple; see Eqn \ref{eqn: 3lvl}) and recombination cooling (green) per unit mass. Displayed optically thin heating expected from the incoming sub-stellar and globally averaged XUV flux of (see Eqn \ref{eqn: heat_thin}). In blue is the cooling from the auroral transition of NI and in red is the cooling effect of mixing in $10\%$ oxygen,  which is dominated by the nebular transition. Lastly, in green is the cooling from radiative recombination of N$^+$. (A) young Sun Mars $(n_0, \lambda_0, \gamma_{\scriptscriptstyle \mathrm{on}})=$ (\SI{e14}{\per\cm\cubed}, 105, 0.826). Cooling is calculated by assuming an upper-bound uniform free electron density of \SI{e6}{\per\cm\cubed} and corresponding density of N$^+$ (NII) by quasineutrality. (B) Cooling for the young Sun Earth hydrodynamic onset is calculated using electron profile from \cite{Johnstone2019}, which peaks at $10^{6.5}$ electrons \unit{\per\cm\cubed}. The effect of doubling the estimated electron densities on the total line cooling from nitrogen is shown in dashed-blue ($q_{cool}(2\times e^{-})$).}
\end{figure*}

Exploring [NI] in Figure \ref{fig: grotrian}, the long radiative lifetime of the first excited level at \SI{2.4}{\electronvolt} above ground (indexed 2) of $\mathrm{NI}$ at \SI{21}{hours} means that the $2({}^2 D)  \to 1({}^4 S)$ `nebular’ forbidden transition is a poor cooling mechanism. The second excited level, or third level of the atom, contributes two lines at $E_{31}=$ \SI{3.6}{\electronvolt}: the `transauroral’ transition $3({}^2 P)  \to 1({}^4 S)$ has transition rate $A_{31} =$ \SI{5.22e-3}{\per\second},
which is minor compared to the $3({}^2 P) \to 2({}^2 D)$ `auroral’ transition with $A_{32}=$ \SI{8.47e-2}{\per\second}. Thus, the third excited level has a lifetime of \SI{12}{\second} and cooling is dominated from the first three levels by $\left[\mathrm{NI}\right]_{1.04\mathrm{\mu m}}^{3 \to 2}$ with  $E_{32} =$ \SI{1.2}{\electronvolt}. 

We use a cool-to-space approximation as the forbidden lines make up a small optical depth due to their small Einstein-A \citep{NLTE}. We calculate the optical depth assuming Doppler broadening only: 
\begin{equation}
    \tau_{\nu_{ul}} =  \left (\frac{h c_{l}}{E_{ul}} \right)^3 \frac{\sqrt{\gamma} g_u N}{8 \pi^{3/2} g_l} \frac{A_{ul}}{c_{\gamma}}, 
\end{equation}
where $N$ is the total column density above the thermobase, $h$ Planck's constant and $c_{l}$ the speed of light  \citep{Kunc1989}. The escape probability formulation used in \cite{Nakayama_2022} is not necessary for this regime as line photons propagated below the thermobase are lost as blackbody radiation - still contributing a net cooling to the thermosphere.

\subsection{Is Line cooling Minor in the Weakly ionized Onset of Hydrodynamic Escape?}
\label{subsec: LC}
\subsubsection{Young Sun Mars}
\label{subsec: YSM-LC}

The heating at the sonic point in the polytropic onset model for Young Sun Mars $(\SI{e14}{\per\cm\cubed}, 105, 0.826)$ is consistent to within a factor of two of the optically thin heating expected from the incoming globally averaged XUV flux of \SI{0.1}{\watt\per \m\squared} (see Eqns \ref{eqn: heat_sonic}-\ref{eqn: heat_thin}). Figure \ref{fig: marsHC} shows how in an advection-dominated nitrogen flow at these temperatures, atomic line cooling is not significant compared to the heating required for model Knudsen onset. We assume an upper-bound uniform free electron density of \SI{e6}{\per\cm\cubed}. NI is inefficient at cooling compared to CI and OI at these temperatures. For example, the Einstein coefficient for the $2 \to 1$ transition is approximately $700\times$ larger for OI than NI, explaining how a 10\% mixing ratio of oxygen would dominate the cooling. However, the model heating is more than $10\times$ the OI cooling, so we would still expect consistent onset hydrodynamic escape even for a pure oxygen thermosphere. Radiative recombination (see Section \ref{sec: thermostat}) contributes the greatest cooling below \SI{2000}{\km}.
Mars, with a gravitational binding atmosphere a fifth of Earth's, thus represents well the limit where line cooling is not the controlling factor in the retention of a secondary atmosphere.

\subsubsection{Young Sun Earth}
\label{subsec: YSE-LC}
The Knudsen-onset from the polytopic model for a nitrogen-dominated atmosphere $(\SI{e15}{\per\cm\cubed}, \ 416, \ 0.77)$ is again within a factor of two of the optically thin heating expected from the incoming globally averaged XUV flux of \SI{0.8}{\watt\per \m\squared} found to be required in Section \ref{subsec: YSE}. However, compared to Mars, the hotter aeronomy required to escape Earth’s gravitational well results in significant line cooling. The electron density profile used for all the curves is from \cite{Johnstone2019}, where we also double the electron density in the profile as a sensitivity test. The profile has a peak electron density of \SI{3e6}{\per\cm\cubed}, above which the density decreases towards the sonic point and beyond. 

The contributions to cooling are as follows. The blue curves in Figure \ref{fig: YSEHC} show cooling from the auroral transition of neutral nitrogen. In addition, the cooling effect from mixing in $10 \% $ atomic oxygen is shown through the dominant nebular transition of OI in red. Figure \ref{fig: YSEHC} shows how cooling is minor at the sonic point compared to the XUV heating. The auroral cooling of NI at \SI{1040}{\nano\metre} in blue almost matches the implicit heating in the middle atmosphere. The NII cooling assumes nitrogen ion density equal to the electron density by quasineutrality. Due to shorter radiative lifetimes (see Fig \ref{fig: grotrian}), the line cooling of NII becomes dominant at the sonic point despite being a minor fraction. The OI nebular line dominates at the bottom of the atmosphere, where temperatures are low. Radiative recombination of atomic ions and electrons is minor in the upper atmosphere. 

Though the cooling integrated over the model atmosphere is minor, the comparable heating and cooling in the middle atmosphere leads us to set the efficiency sub-factor from cooling as $\eta_c = 0.5$. So, overall efficiency of the outflow $ \eta \approx \eta_{\gamma} \eta_{\mathrm{pr}} \eta_{c} \approx 0.13$. An XUV flux twice as intense, corresponding to $\sim 400 F_{\scriptscriptstyle \mathrm{XUV}, \earth}$, is then enough to account for non-LTE cooling. When the  ionization fraction is doubled compared to \cite{Johnstone2019} the total nitrogen cooling dominates over the $\sim 200 F_{\scriptscriptstyle \mathrm{XUV}, \earth}$ polytropic heating in the atmosphere above \SI{3000}{\km} (Fig  \ref{fig: YSEHC}). To explore these sensitivities, we will next explore the dynamics of hydrodynamic escape and hydrostatic instability with freely varying electron densities in an analytic model of the collisional-radiative thermostat. 
\section{collisional-radiative Thermostat and Global Ion Outflow} \label{sec: thermostat}
We now focus our attention towards the dynamics of predominantly ionized outflow. We show that the large escape velocity of super-Earths means that an XUV-driven global ion outflow is the only route towards rapid escape of a secondary atmosphere. The formulation of ion-electron effects in the instability of forced hydrostatic atmospheres is also key to understanding the differing conclusion of the present study from that of \cite{Nakayama_2022}. 

\subsection{Plasma Escape Temperature}
\label{subsec: plasma}
Higher ionization fractions lead to greater line cooling, reducing the tendency of an atmosphere to escape. The counter-effect, hitherto unremarked upon in the context of global loss of secondary atmospheres, is the increase in sound speed in an ionized atmosphere, or equivalently, the reduction in effective mean molecular weight, and thus a reduced escape temperature via $GM\mu /( k_{B} T_{esc} r)=2$ (see Eqn \ref{eqn: Tesc}). An ambipolar electrostatic field couples the plasma, reducing escape of electrons and promoting escape of ions \citep[e.g.,][]{Bauer&Lammer, Koskinen}.

The ion-acoustic speed of sound for a quasineutral mixture without secondary ionization is 
\begin{equation}
c_{i-a}^2 = \frac{ \gamma_e f_{+} k_B T_e + \gamma_i f_{+} k_B T_i + (1 - f_{+}) \gamma_n k_B T_n}{f_{+}m_e + f_{+}m_i + (1-f_{+})m_n},
\label{eqn: ion-acoustic}
\end{equation}
where $f_{+}$ is the ionization fraction, and $m_e, m_i$ and $m_n$ refer to the electron, ion and neutral atom masses respectively \citep[e.g.,][]{ChenPlasma}. We can assume ion-electron thermalization and approximately isothermal outflow in these high XUV flux conditions, such that neutral, ion and electron temperatures and ratios of specific heat are equal $T_i \approx T_e, \gamma_i =\gamma_e =1$. The ion-acoustic sound speed can then be simplified to
\begin{equation}
 c_{i-a}^2\approx \frac{ (1 + f_{+}) k_B T_i}{m_i}.
\end{equation}
 Thus, the mean molecular weight will be given by 
\begin{equation}
    \mu(f_{+}) \approx \frac{m_{i} }{1 + f_{+}}, \label{eqn: mmw}
\end{equation}
so for $f_{+}=0.5$ or $1$, the mean molecular weight is $2/3$ or $1/2$ the neutral equivalent. We define the local \textit{plasma} escape temperature as the escape temperature when the local ionization fraction is roughly unity:
\begin{equation}
    T_{\mathrm{esc}}(f_{+}=1) = \frac{m_{i}}{2}\frac{v_{\mathrm{esc}}(r)^2 }{4 k_{B}} = \frac{T_{\mathrm{esc}}(f_{+}=0)}{2}. \label{eqn: TescPlasma}
\end{equation}
As displayed in Figure \ref{fig: thermostat}, the escape temperature for neutral outflow from Earth is approximately \SI{18000}{\kelvin} but for a fully ionized atmosphere only \SI{9000}{\kelvin}. This simple derivation captures the principal physics, but ionospheric processes are non-linear and this effect could also be enhanced \citep{EarthField}. 
\subsection{Ion-Electron Recombination}
In the lower and upper layers of Earth's ionosphere, the primary sinks of photoelectrons are dissociative recombination of molecular ions and ambipolar diffusion, respectively \citep{Bauer&Lammer}. These mechanisms require a background of molecular and neutral species, which close the chemical network via charge transfer. The radiative recombination of ions and electrons, though significant in the F2 layer, remains minor throughout the atmosphere \citep{Solomon_2010}. However, under the raised XUV fluxes that yield extended ionospheres and greater total electron content, radiative recombination can become the primary sink for photoelectrons \citep[e.g.,][]{Nakayama_2022}. 

To introduce the role of recombination, we first compare the dynamics of radiative-recombination cooling with atomic line cooling. The rate of dielectronic recombination, which proceeds via a metastable intermediate, was calculated with a fit from \cite{Zatsarinny2004, Erratum} and found to be minor for \SI{200}{\kelvin} $<T<$ \SI{20000}{\kelvin}, only matching the radiative recombination rate at $\approx$ \SI{27000}{\kelvin}. Hence, dielectronic recombination is negligible for the conditions considered in this study. The derivation of radiative recombination (RR) rates takes advantage of photoionization cross-section data, being the time-reversed process, and is fit with the following formula from \cite{Badnell_2006}:  
\begin{multline}
    \alpha_{\mathrm{RR}}(T) = A \left[\sqrt{T / T_0}\left(1+\sqrt{T / T_0}\right)^{1-B-C\exp(-T/T_2)} \right. \\ \left . +\left(1+\sqrt{T / T_1}\right)^{1-B-C\exp(-T/T_2)} \right ]^{-1},  
\end{multline}
where $(T_0, T_1, T_2, A, B, C) =$ (\SI{9.467e-2}{K}, \SI{2.954e6}{K}, \SI{6.379e4}{K}, \SI{6.387e-10}{\cm \cubed \per \second}, 0.7308, 0.2440). The timescale of thermalization via collisions $t_{th}$ remains shorter than the recombination timescale $t_{RR}$ even when the photoionization rate $1/t_{pi}$ is fast, $t_{th} \ll t_{RR} \sim t_{pi}$, meaning that for each creation of a photoelectron, a \textit{thermal} electron and an ion recombine. The average photon energy from recombination to N is $\sim h \nu_{\mathrm{I}} + 3 k_{B} T_{e}/2$, but the thermal loss from the electron gas is only the continuum part $Q_{RR} \propto 3k_{B}T_{e}/2$. In radiative recombination, cooling is less than total emission because the potential energy an atom gains during ionization is retained, rather than being transferred to the electron gas. Accordingly, the ionization potential is considered as an inefficiency of photoionization heating $\eta_{pi} \approx 1 -  \nu_{\mathrm{I}}/ \overline{\nu}$ throughout our calculations. An exact formula for recombination cooling can be found in \cite{Tucker66}. 

\begin{figure}[htpb!]
\centering
\includegraphics[scale=0.36]{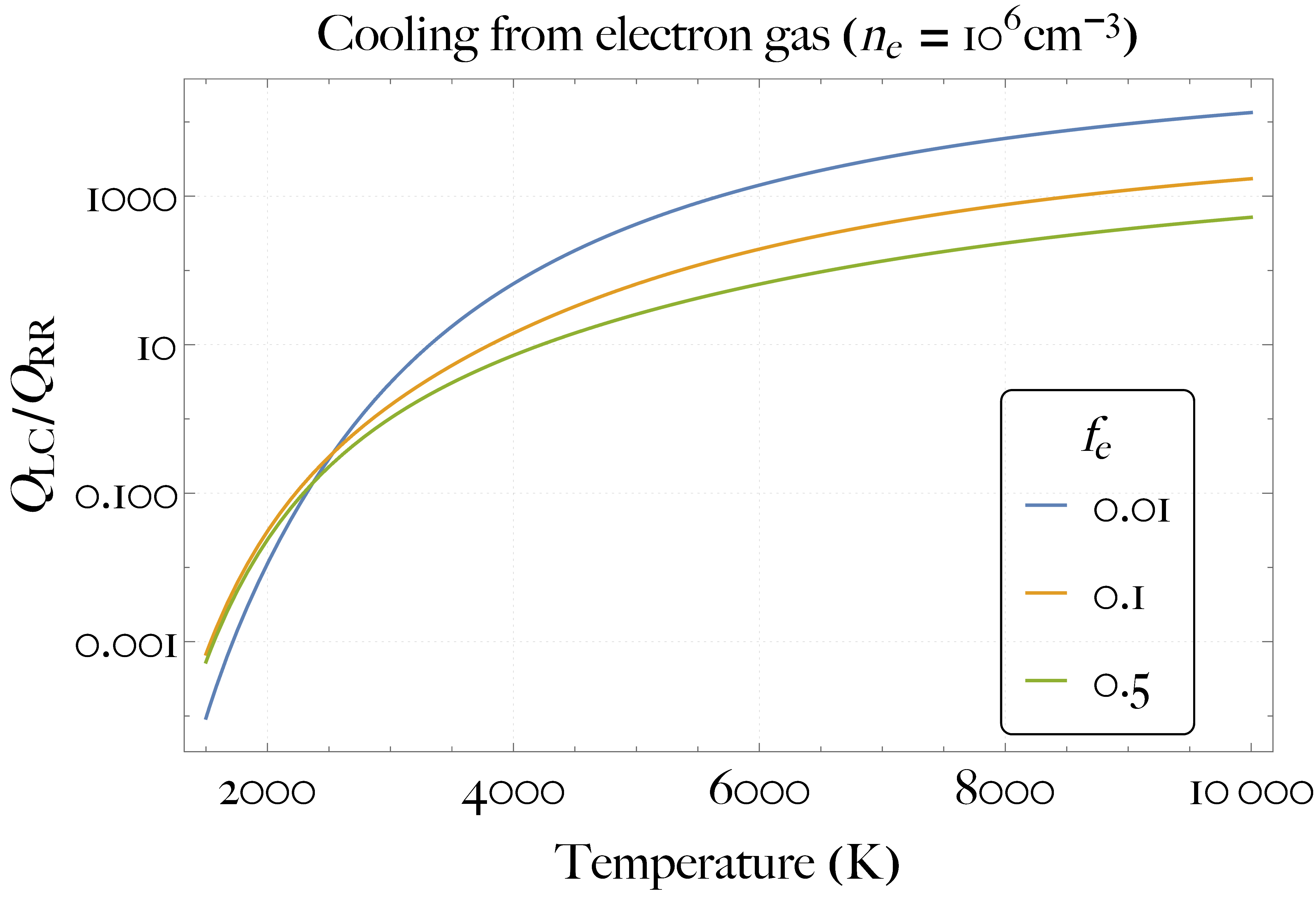}
\caption{Ratio of local cooling luminosities from excitation of atomic lines $Q_{LC}$ and ion-electron radiative recombination $Q_{RR}$. The electron density is \SI{e6}{\per\cm\cubed}. The density of neutral nitrogen atoms is varied from \SI{e8}{\per\cm\cubed} (blue) and \SI{e7}{\per\cm\cubed} (orange) and \SI{e6}{\per\cm\cubed} (green) at temperatures from \qtyrange{1500}{10000}{\kelvin}. This is roughly equivalent to ionization fractions of $f_+ = 0.01, 0.1, 0.5$. \label{fig: RRLC}} 
\end{figure}

We compare the ratio of the volumetric line cooling to RR cooling in Figure \ref{fig: RRLC} for varying ionization fractions $f_{+}$. Recombination is dominant by several orders of magnitude at $\sim$\SI{e3}{\kelvin}, with excitation of atomic lines increasing steeply with temperature until matching recombination at $\sim$\SI{3000}{\kelvin} and going on to dominate by up to several orders of magnitude at \SI{e4}{\kelvin}. \cite{Nakayama_2022} calculate radiative recombination via an unspecified prescription and do not discuss its dynamics in detail, so direct comparison is unavailable. However, their conclusion that the key role of forbidden-line cooling in escape is not superseded by recombination cooling agrees with the work reported here. 

\subsection{The Thermostat of Photoionization versus Line Cooling} \label{subsec: thermostat}

\begin{figure*}
\centering
\includegraphics[scale=0.65]{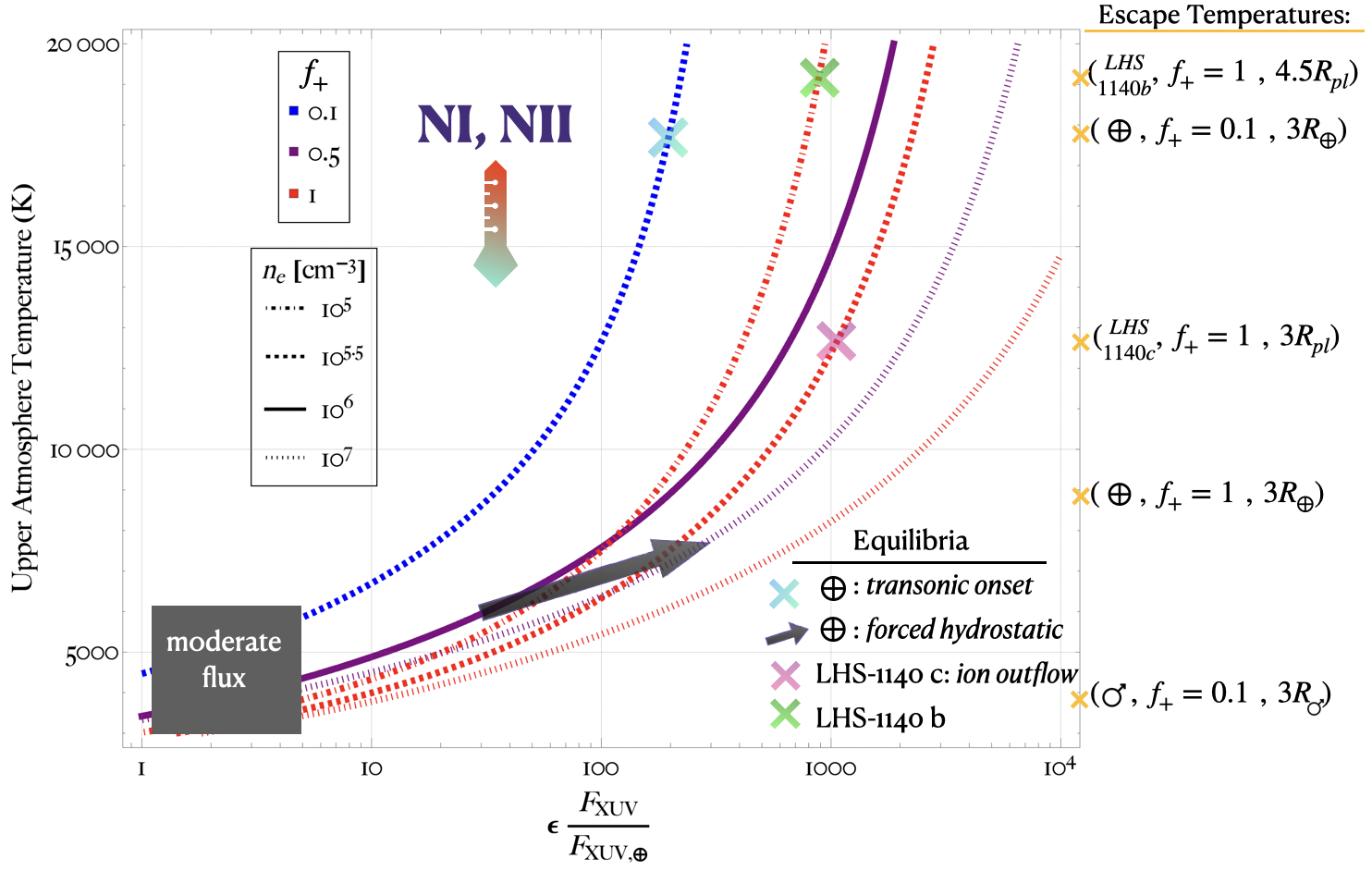}
\caption{\label{fig: thermostat} The upper atmosphere temperatures determined locally by the collisional-radiative thermostat as a function of XUV flux relative to Modern Solar conditions and electron densities. The escape temperatures are shown for each planet with a range of ionization fractions $f_+$ (Eqns \ref{eqn: Tesc} \& \ref{eqn: mmw}). \textit{Note for interpretation that a hydrostatic exobase at greater than half the escape temperature is unstable and the sonic point of a hydrodynamic outflow will approximately reach the escape temperature.} A range of ionization fractions $0.1-1$ and local electron densities \SI{e5}{} -- \SI{e7}{\per\cm\cubed} are identified by legends on the lefthand side. Advection reduces the energy available from XUV heating, along with thermal conduction and the assumed minor optical attenuation; expressed in the prefactor $\epsilon$. The blue cross highlights a local equilibrium at the sonic point that, when accounting for advection through $\epsilon$, is roughly consistent with the Young Sun Earth polytropic onset at $\sim 400 F_{\scriptscriptstyle \mathrm{XUV}, \earth}$ (see Section \ref{subsec: YSE}). The black arrow illustrates a feedback between photoionization and line cooling on temperature relevant to \cite{Nakayama_2022}. However, if the hydrostatic Earth-like atmosphere reaches high ionization fractions in the black-arrow evolution it becomes unstable to escape. The pink and green crosses roughly indicate the XUV fluxes required to drive global ion outflow from LHS 1140 c and LHS  1140 b based on their plasma escape temperatures. Exoplanet data are taken from \cite{Cadieux_2024}. Transonic mass loss rates are proportional to sonic-point densities (Eqn \ref{eqn: interpolate}).}  
\end{figure*}

In this sub-section, we explore how photoionization both heats the atmosphere but also generates free electrons that excite greater line cooling. We consider quasineutral nitrogen layers in the upper atmosphere that are optically thin to incoming XUV photons. The rate ions are advected away, through either acceleration in bulk outflow or molecular diffusion, can be included such that the transport equation for N becomes 
\begin{equation}
 \frac{F_{\scriptscriptstyle \mathrm{XUV}} \overline{\sigma_{\nu_{\mathrm{I}}}}}{h\overline{\nu_{\mathrm{I}}} } n_{\mathrm{N}} \approx n_{e} \alpha_{\mathrm{RR}} n_{\mathrm{N^{+}}} + nu \dv{f_{+}}{r}, \label{eqn: ion-sim-recom}
\end{equation}
where $f_{+} = \frac{n_{\mathrm{N^{+}}}}{n}$ and we have implemented a monochromatic approximation at the front \citep[e.g.,][]{Murray_Clay_2009}. The mean of energy of photons producing singly and doubly ionized nitrogen is $h \overline{\nu_{\mathrm{I}}} \approx $ \SI{33.6}{\electronvolt} and $h \overline{\nu_{\mathrm{II}}} \approx $ \SI{46.5}{\electronvolt}, and the mean absorption cross section are similar: $\overline{\sigma_{\nu_{\mathrm{I}}}}\approx \overline{\sigma_{\nu_{\mathrm{II}}}} \approx 10^{-17}$ \unit{\cm\squared} evaluated for a Modern Active Sun Spectrum \citep{Huebner2015}.

The cooling from the forbidden line cooling of NI and NII must balance the net heating from photoionization: 
\begin{multline}
       n \epsilon F_{\scriptscriptstyle \mathrm{XUV}}^{\theta} \left( f_{+} \frac{ \Delta \nu_{\mathrm{II}}} {\overline{\nu_{\mathrm{II}}} } \overline{\sigma_{\nu_{\mathrm{II}}}} +(1-f_{+}) \frac{ \Delta \nu_{\mathrm{I}}} {\overline{\nu_{\mathrm{I}}} } \overline{\sigma_{\nu_{\mathrm{I}}}} \right) \approx \\ \sum_{\mathrm{NI}}^{\mathrm{NII}}  \sum_{j,i>j}^{1\to3} n_{i}(T,n_e) A_{ij} E_{ij} \ ,
    \label{eqn: thermostat} 
\end{multline}
where the average excess energies $\Delta \nu_{\mathrm{I,II}} = \overline{\nu_{\mathrm{I,II}}} - 3k_{B} T_{e}/2 - h \nu_{\mathrm{I,II}}$ and the number densities $n_i(T,n_e)$ are solutions from the three-level model (Eqn \ref{eqn: 3lvl}). The efficiency prefactor $\epsilon$ is the fraction of local heating sunk into line cooling, as opposed to advection $Q_{\mathrm{adv}}$ and conduction $Q_{\mathrm{cond}}$. Minor optical attenuation is also represented in $\epsilon \approx \mathrm{e}^{-\tau}(1 - (Q_{\mathrm{cond}}+Q_{\mathrm{adv}})/Q_{\scriptscriptstyle \mathrm{XUV}})$, where $\tau \lesssim 1$. Even if advection is major for the transport of ions, it may have a minor contribution to the energy balance. Thermal conduction down the thermospheric inversion can contribute significant cooling locally, but the front may also be found where the temperature profile has flattened. Thus, $\epsilon$ takes values $\sim 0.1 - 1$.

The collisional-radiative thermostat effect of Equation (\ref{eqn: thermostat}) is illustrated in Figure \ref{fig: thermostat}. For fixed electron density, order of magnitude changes in the XUV flux cause order unity changes in the temperature generated. However, higher XUV flux results in a greater electron density in the front, which increases line cooling and feedbacks on the temperature; see forced-hydrostatic evolution (black arrow, Fig \ref{fig: thermostat}). Optically thin absorption is proportional to the electron density (via the total density), while the line cooling dependence has a proportionality closer to the square of the electron density, which is exact in the two-level coronal model (Eqn \ref{eqn: 2lvl}). Increasing electron density has a more marginal effect when approaching the critical density for LTE (see Sec \ref{subsec: LC}), which is sometimes referred to as quenching \citep{Bauer&Lammer}. For the transonic outflow equilibria, the mass loss rate can be interpolated from Figure \ref{fig: thermostat} using
\begin{equation}
    \Phi_{hyd} \simeq 4 \pi r_{sc}^2 (m_i n_e/f_+) \sqrt{k_{B} T_{sc} / \mu} \ . \label{eqn: interpolate}
\end{equation}

\begin{deluxetable*}{cccccccc}
\tablecaption{Equilibria of Photoevaporating Nitrogen Atmospheres \label{tab: equilibria}}
\tablewidth{0pt}
\tablehead{
\colhead{\textit{Transonic:}} & \colhead{$\times F_{\scriptscriptstyle \mathrm{XUV}, \earth}$} & \colhead{Vertical eqm} & \colhead{Primary balance} & \colhead{$f_+$} & \colhead{Consistent} & \colhead{Comment} & \colhead{ $\Phi_{\mathrm{hyd}}(\mathrm{bar}/\mathrm{MYr})$}}
\renewcommand{\arraystretch}{1.2}
\startdata
YS Mars & $50$ & hydrodynamic & ionization-advection &  $\lesssim 0.1$ &  \checkmark & $q_{\gamma} \gg q_{c}$, EL & $2$ \\ \hline
\multirow{2}{*}{YS Earth} & \multirow{2}{*}{400} & hydrostatic & ionization-recombination & $\lesssim 1$ & $\times$ & $\to$ ion outflow \\ 
& & hydrodynamic & ionization-advection & $\sim 0.1$ & \checkmark &$q_{\gamma} \gtrsim q_{c}$ & $6$ \\ \hline
LHS 1140 c  & $1000$ & hydrodynamic & ioniz.-adv.-recom. & $\sim 1$ & \checkmark ($?$) & $1.9 M_{\earth}$, CRT & $\sim 1.2$ \\ \hline
LHS 1140 b  & $2000$ & hydrodynamic & ioniz.-adv.-recom. & $\sim 1$ & \checkmark ($?$)  & $5.6 M_{\earth}$, CRT & $\sim 0.9$ \\ 
\enddata
\tablecomments{ YS stands for the Young Sun scenarios modeled. $q_{\gamma}$ and $q_{c}$ refer to local XUV-heating and atomic line cooling with altitude. EL refers to following the energy-limited behavior with the caveat of the onset and CRT refers to the onset being determined from the collisional-radiative thermostat rather than an adjusted energy limit. The transition to ($\to$) ion outflow refers to transient evolution from an unstable hydrostatic atmosphere. The abbreviation ioniz.-adv.-recom. refers to ionization balanced by advection and recombination as for the steady state of global ion outflow. Numbers are based on the analytic modeling reported in the present study. This table of best estimates requires follow-up work with hydrocodes and molecular-kinetics simulations. In particular, metal outflows from super-Earths are poorly understood and here we make only tentative suggestions; as indicated by the question marks.}
\end{deluxetable*}

\subsection{\texorpdfstring{Comparison to \cite{Nakayama_2022}}{Comparison to Nakayama 2022}}
\label{subsec: Nakayama}
The collisional-radiative thermostat (Eqn \ref{eqn: thermostat}) offers a broad physical explanation of the surprisingly uniform temperature dynamics of the exobase at high XUV fluxes in \cite{Nakayama_2022}: a feedback between forbidden line cooling and photoionization heating mediated by electron density.
\cite{Nakayama_2022} calculate that for $500 F_{\scriptscriptstyle \mathrm{XUV}, \earth}$ the ionization front $f_{+}=0.5$ occurs at an altitude of $\sim$ \SI{3000}{\kilo \metre}, reaching a peak temperature of \SI{5500}{\kelvin}. The ionization front temperatures for $50 \to 1000 F_{\scriptscriptstyle \mathrm{XUV}, \earth}$ are within a \SI{500}{\kelvin} temperature interval, indicating a thermostatic effect must be present. In Figure \ref{fig: thermostat}, we explore a reduced model of the ionization with the forced hydrostatic arrow taking the range $(\epsilon F_{\scriptscriptstyle \mathrm{XUV}}, \ n_e, \ f_e): (30, \ 10^6 \ \mathrm{cm}^{-3}, \ 0.5) \to (300, \ 10^7 \  \mathrm{cm}^{-3}, \ 0.5)$. Equation \ref{eqn: thermostat} yields $T_{stat}: 6000 \to 8000 \ \mathrm{K}$ and above the front ionization fractions reach unity. Compared to \cite{Nakayama_2022}, the excess temperature may be explained through their unspecified treatment of recombination cooling and through our neglecting of oxygen line cooling for this figure.  

However, the present study does not concur with \cite{Nakayama_2022} that the thermostat allows survival of an Earth-like atmosphere up to $1000 F_{\scriptscriptstyle \mathrm{XUV}, \earth}$. Though the forced-hydrostatic Earth-like equilibrium in Figure \ref{fig: thermostat} remains at less than half of the fixed-neutral escape temperature, it is the plasma escape temperature that should be considered for ionization fractions close to unity (Eqn \ref{eqn: TescPlasma}). The forced-hydrostatic evolution in Figure \ref{fig: thermostat} exceeds half the plasma escape temperature, almost reaching it, and would outflow. In general, the intermediate free evolution from hydrostatic instability to hydrodynamic escape could occur through the fluid expansion of ions and electrons aloft of a neutral N-exobase, which drags the atmosphere through the exobase in a drifting Maxwellian \citep{Volkov2011b}. 

We also highlight the apparent lack of sensitivity of the height of the model exobase to increasing XUV flux in \cite{Nakayama_2022}. On the flux increasing by a factor of ten from $50 \to 500 F_{\scriptscriptstyle \mathrm{XUV}, \earth}$, the apparent height increases only by 20\%, or roughly a single local scale height. We expect that the increase of peak electron density with XUV flux should push the exobase significantly further out because (1) the scale height increases with the reduction of mean molecular weight from ambipolar diffusion and (2) the more frequent electron-neutral and ion-neutral collisions allow the neutral-N exobase to form at orders of magnitude lower densities. Moreover, at $500 F_{\scriptscriptstyle \mathrm{XUV}, \earth}$ \cite{Nakayama_2022} find the ionization fraction at the top of the atmosphere to be unity, so that before the mean free path of nitrogen atoms grows large enough for ballistic escape, the atmosphere is ionized enough that formation of a neutral exobase is precluded. 

The polytropic model and the transonic onset equilibrium for Young-Sun Earth in Figure \ref{fig: thermostat} roughly agree that $400 F_{\scriptscriptstyle \mathrm{XUV}, \earth}$ provides enough heating to drive escape at the neutral onset as a conservative estimate. So, together with the instability of the forced hydrostatic equilibrium (Fig \ref{fig: thermostat}) relative to the plasma escape temperature, our idealized modeling offers a consistent picture. 
\subsection{TRAPPIST-1 b: Airless}
\label{subsec: T1b}
TRAPPIST-1 b has radius $1.16R_{\earth}$ and mass $1.37M_{\earth}$ \citep{Agol_2021}. Its escape velocity is then \SI{12.2}{\km\per\second}, less than 10 \% different from Earth, so we expect a similar escape regime for similar XUV fluxes. We do not plot TRAPPIST-1 b's escape temperatures in Figure \ref{fig: thermostat} because they are similar to Earth's. 

At significantly higher XUV fluxes than Young Sun Earth, the dominant balance of photoionization is expected to be from radiative recombination as opposed to advection; this has been considered in the photoevaporation of hydrogen envelopes \citep[e.g.,][]{Murray_Clay_2009, Owen2012}. Furthermore, the transonic mass loss rate is not expected to rise in proportion to XUV flux indefinitely due to the limited penetration of XUV heating and the increasing severity of temperature inversion required against the effects of conduction and line cooling. To calculate the XUV flux for which mass loss rates plateau would require a hydrocode with the correct photochemistry and electrodynamics, so is beyond the scope of this work. However, a hypothetical atmosphere of nitrogen could outflow from TRAPPIST-1 b at $(\epsilon F_{\scriptscriptstyle \mathrm{XUV}}, \ n, \ f_+): (10^{3.5}, \ 10^7 \mathrm{cm}^{-3}, \ 1)$ (Fig \ref{fig: thermostat}) with a close-in sonic point, possibly a detached D-type shock \citep{Spitzer}.

Thus, unless TRAPPIST-1 b formed extremely volatile rich, it is unlikely that any atmosphere would remain today. Indeed, \cite{Van_Looveren_2024} calculate that even the present-day fluxes of ionizing radiation in the TRAPPIST-1 planetary system are intense enough to preclude the retention of any Earth-like or Venus-like atmosphere. However, we note that they use a forced-hydrostatic Jeans prescription of escape that does not include the key electric-dipole-forbidden lines.
\subsection{LHS 1140 b and c: Airy?}
The escape temperature for a predominantly neutral flow from LHS 1140 c ($1.9M_{\earth}$, $1.27R_{\earth}$,  0.027 AU) would be $ \gtrsim $ \SI{22000}{\kelvin} due to the $1.5 \times$ tighter gravitational binding that of Earth. However, temperatures would be kept lower by the collisional-radiative thermostat under rising XUV fluxes even for greater than $1000 F_{\scriptscriptstyle \mathrm{XUV}, \earth}$, so transonic escape with a significant neutral fraction is prevented. The ion-acoustic speed in an ion outflow allows the plasma escape temperature to be half as large as the neutral. The pink cross in Figure \ref{fig: thermostat} explores a hypothetical sonic point with an electron density of $10^{5.5}$ cm$^{-3}$ requiring an XUV flux of $10^3 F_{\scriptscriptstyle \mathrm{XUV}, \earth}$ to drive escape of one bar every million years. We estimate $\epsilon \sim 1$, assuming that advection is only a minor sink at this flux and the terminal profile to be relatively isothermal. Thus, escape would not be driven for LHS 1140 c with its contemporary instellation of $10^2 F_{\scriptscriptstyle \mathrm{XUV}, \earth}$ \citep{Spinelli_2023}, but a secondary atmosphere formed in the early pre-main-sequence with $\gtrsim 10^3 \times F_{\scriptscriptstyle \mathrm{XUV}, \earth}$ could well have been lost. 

The companion planet LHS 1140 b ($5.6M_{\earth}$, $1.7R_{\earth}$,  0.09 AU) has a $3\times$ deeper gravitational well than Earth, so the neutral escape temperature would be in the ballpark of $\gtrsim$\SI{50000}{\kelvin}. As a result, only an ion outflow is feasible for the hydrodynamic escape of a secondary atmosphere from a `radius valley' planet. Inspecting the green cross in Figure \ref{fig: thermostat}, even if the sonic point could form as far out as $4.5R_{pl}$, then assuming $\epsilon\sim 0.5$ and an electron density $10^{5}$ cm$^{-3}$ an XUV flux of $2000 F_{\scriptscriptstyle \mathrm{XUV}, \earth}$ would be required. Since temperature must rise steeply to reach $\sim 2\times 10^4$ K, thermal conduction is expected to be significant and we estimate $\epsilon \sim 0.5$. If escape at lower densities is possible, the required flux would be reduced but so would the escape rate (Eqn \ref{eqn: interpolate}). In the LHS 1140 system, planet b only receives one twelth of the instellation of c. So, a thick secondary atmosphere is unlikely to be photoevaporated even accounting for early pre-main-sequence fluxes.

Whether LHS 1140 c would retain a thick secondary atmosphere is sensitive to its initial conditions, but LHS 1140 b should have retained its atmosphere, consistent with observations from \cite{Cadieux_2024}. For cool rocky planets orbiting low-mass stars, the higher escape temperatures of super-Earths compared to sub-Earths is crucial to prospects for the retention of secondary atmospheres. 
\section{Discussion}
\label{sec: discuss}
\begin{figure*}
\centering
\includegraphics[scale=0.5]{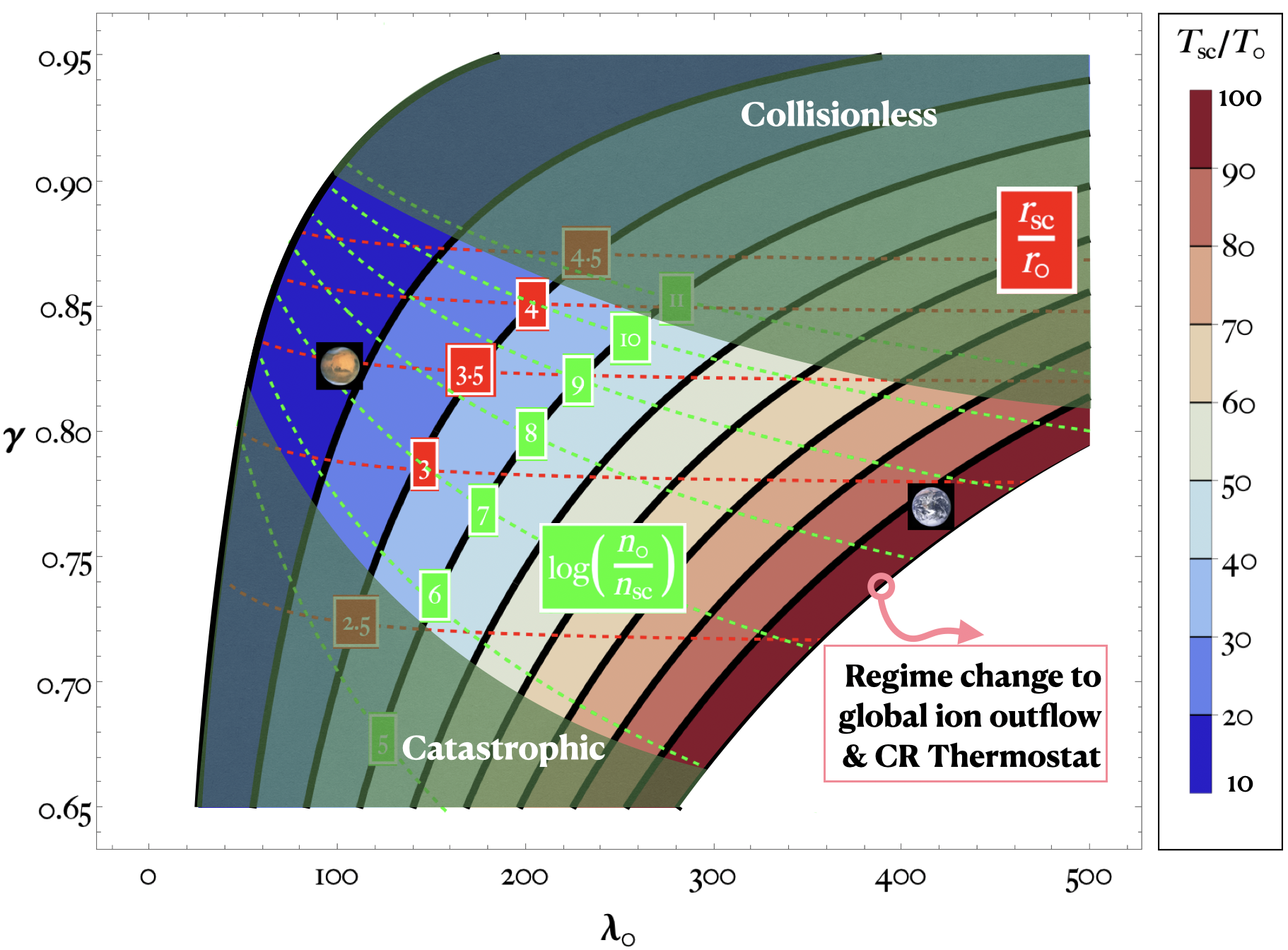}
\caption{Regime diagram for high molecular weight escape in terms of the polytropic index $0.65 \leq \gamma \leq 0.95$  and the hydrodynamic escape parameter $20 \leq \lambda_{0} \leq 500$. The heat map indicates the ratio of sonic to base temperatures $10-100$ (Eqn \ref{eqn: Tsonic}). Contours of uniform ratio of sonic to base radius (red-dashed, Eqn \ref{eqn: itr}) and uniform sonic to base density in log units (green dashed, Eqn \ref{eqn: rho-sc}). The Young Sun Earth (c.f. Figures \ref{fig: YSE-structure} \& \ref{fig: YSEHC}) and Young Sun Mars (see Figures \ref{fig: marsT} \& \ref{fig: marsHC}) calculations are represented by images of each planet. This figure can be compared to the solution space of polytropic solutions (Fig \ref{fig: SolSpace}). For large enough $\lambda_0$, the polytropic model of steep XUV inversions breaks down because line cooling will begin to dominate over advection. The overlayed pink patch is to illustrate that the shift is broad and dependent on boundary conditions and dimensional parameters. Rightward of the transition, the collisional-radiative thermostat will control the resulting global ion outflow (see Section \ref{sec: thermostat}).} \label{fig: conceit}
\end{figure*}

The rapid escape of heavier species is usually only considered in drag-off by hydrogen, so the direct hydrodynamic escape of a nitrogen or carbon dioxide atmosphere is often left out of atmospheric evolution studies of rocky planets \citep[e.g.,][]{Teixeira_2024}. In fact, for a planet with a thick secondary atmosphere, such as Venus, other erosion mechanisms are generally too slow or variable, leaving direct hydrodynamic escape as the main route to airlessness \citep[e.g.,][]{Gronoff}. Early \textit{JWST} observations are revealing a pattern of cool rocky worlds that are airless or thinly blanketed. Thus, observations and theory suggest that atmospheric retention could be a major bottleneck in the occurrence of observable Earth-like habitats $\eta_{\earth}$ \citep[e.g.,][]{scherf2022}.
\subsection{Landscape of High Molecular Weight Escape}
The polytropic framework allows a comparative planetology perspective, with the resulting insight resolvable into the simplified representation of the range of governing physics. Figure \ref{fig: conceit} has $\lambda_0$ on the horizontal axis: the ratio of gravitational binding to the thermal reservoir at the thermobase. On the vertical axis is $\gamma$ describing the steepness of the temperature inversion, which is implicitly determined by XUV instellation (Eqn \ref{eqn: kin_poly}). Ratios of base to sonic densities, temperatures, and radii can be read from Figure \ref{fig: conceit}. The diagram can be used as a visual aid to see for which star-planet conditions hydrodynamic escape will be possible.

The pairwise base-to-sonic ratios in log units map to the proportionally varying Knudsen number at the sonic point if holding base density constant.  The right-hand edge where collisional onset may be possible is roughly marked as where the sonic density is ten orders of magnitude smaller than at the base. On the $\log \left [n_{0}/n_{sc}\right]= 10$ contour itself, extremely deep penetration of XUV would be required for the flow to launch the outflow from that dense portion of the atmosphere. The left-hand edge is given by the $\log \left [n_{0}/n_{sc}\right]= 6$ contour, where the sonic Knudsen number would likely be more than an order of magnitude smaller than at onset. In this high XUV-flux limit, the assumption of predominantly neutral outflow is likely to break down. Furthermore, the mass loss will be catastrophically fast.

In the region marked out between $6 < \log \left [n_{0}/n_{sc}\right] < 10$, the sonic radii vary between two and four times the planetary radius. The base to sonic temperature ratio is given by $T_{sc} = \frac{\lambda_0 r_0 T_0}{2 \gamma r_{sc}}$, and is a proxy for when line cooling will become significant in the atmospheres of temperate rocky planets. Accordingly, the consistency of transonic hydrodynamic escape is clear-cut for XUV-evaporation of a thick nitrogen atmosphere on a Mars-like planet due to the lower gravitational binding of the atmosphere. Transonic hydrodynamic escape of the nitrogen atmosphere can be driven with upper atmosphere temperatures staying below $4000\mathrm{K}$, low enough for negligible line cooling. The threshold XUV flux for collisional onset for Mars is calculated as $100\times$ Modern Day Mars XUV, which we highlight as overly conservative compared to previous studies \citep[e.g.,][]{TianMars}, and would entail a rapid loss rate of 20 bars per million years. For much lower XUV fluxes, the threshold for instability of a hydrostatic nitrogen atmosphere at $\sim$ \SI{2000}{\kelvin} would be exceeded.  

For the gravitational binding of the Earth’s atmosphere and a cool thermobase, the temperatures monotonically increased to roughly \SI{23000}{\kelvin} at the sonic point at Knudsen neutral onset. The peak of auroral line cooling from $\left[\mathrm{NI}\right]_{1.04\mathrm{\mu m}}^{3 \to 2}$ and $\left[\mathrm{NII}\right]_{611\mathrm{nm}}^{3 \to 2}$ matches the polytropic heating in the middle atmospheres. Accounting for line cooling through an efficiency  sub-factor $\eta_{c}=0.5$ in the energy limit (Eqns \ref{eqn: E-L} and \ref{eqn: efficiency}), we find the hydrodynamic loss of six bars of nitrogen per million years for $400\times$ Modern Day Solar XUV irradiation. To build confidence in this result, we checked agreement from opposite perspective of the line cooling balancing XUV heating with advection being an efficiency factor. We also reanalyzed the hydrostatic thermosphere profiles of \cite{Nakayama_2022}, finding instability at $500F_{\scriptscriptstyle \mathrm{XUV}, \earth}$, possibly lower, while accounting for the effect of ion-electron interactions on the mean molecular weight as captured by the plasma escape temperature (Section \ref{subsec: plasma}).  Earth can be viewed as occupying yet another sweet spot where the energy limit starts to break down, and the collisional-radiative thermostat takes over for the loss of secondary atmospheres. 

The upper-atmosphere temperatures required to rapidly escape planetary atmospheres with gravitational binding greater than Earth's are prohibitive for the case of predominantly neutral outflow. We hypothesize that even higher XUV fluxes can instead lead to global ion outflow, with atomic line cooling controlling the rate of escape and the collisional regime extending to greater altitudes. Accordingly, a steep polytropic inversion from the thermobase is a poor model. This is highlighted with the righthand regime change in Figure \ref{fig: conceit}.

An onset of rapid escape can be found by matching the collisional-radiative thermostated sonic point with the plasma escape temperature, which is more easily achieved due to the twice-as-fast ion-acoustic sound speed (Eqn \ref{eqn: ion-acoustic}). The plasma escape temperature of the super-Earth LHS 1140 c ($1.9M_{\earth}$) at three planetary radii is roughly \SI{13000}{\kelvin}. Sonic point analysis suggests an uncertain mass loss rate of a bar of nitrogen every million years on exposure to $\sim 10^{3} F_{\scriptscriptstyle \mathrm{XUV}, \earth}$ (Figure \ref{fig: thermostat}). However, more detailed modeling is required to characterize the super-Earth regime beyond the result that escape is appreciably more difficult than from Earth and should take the form of an ion outflow. 

Overall, within model uncertainty, Modern Earth’s atmosphere might well survive in the XUV-irradiation environment of the Young Sun, depending on its initial rotation rate. However, if placed in the habitable zone of an active M dwarf, our analysis suggests that the protective effect of the line cooling is overwhelmed, and a thick secondary atmosphere would likely be lost. Thus, from the perspective of escape, no substantial atmospheres are expected on the TRAPPIST-1 planets unless they formed extremely volatile rich. More advanced modeling and further observations will constrain the reality. For super-Earths orbiting M dwarfs, the zone of atmospheric retention is larger, making them the most promising observational targets for detecting features of an Earth-like habitat.
\subsection{Cosmic Shoreline for Rocky Exoplanets}
The emerging dynamics of the work reported here suggests a non-uniform trend for separating airless and airy planets in the space of XUV flux/fluence against a power law in escape velocity. We venture to make some broad predictions for the cosmic shoreline in contrast with \cite{Zahnle_2017}. The nonlinear onset and rapid rates of transonic outflow suggest that the key parameter governing the escape of secondary atmospheres may be the high-activity maxima of the XUV flux during the main sequence, rather than the lifetime fluence:
\begin{enumerate}[label=\Roman*.,noitemsep,topsep=1pt,parsep=2pt,partopsep=1pt]
\item The activity of a star enhances its  X-ray luminosity $\sim 3\times$ above the long-term average at each evolutionary stage for $\sim$ 10\% of the time \citep{JohnstoneFGKM}. So, if the \textit{average} main-sequence XUV flux is greater than one third of the threshold for transonic escape, a secondary atmosphere is unlikely to survive or be revived. 
\item Otherwise, if main sequence XUV fluxes are low enough that a secondary atmosphere would remain hydrostatic throughout, then its survival will depend on an interplay between stellar wind erosion, volcanic activity and impacts \citep{KiteBarnett}.
\item Accretion of an envelope of nebular hydrogen, by weathering some or all of the saturated phase of coronal activity on the pre-main-sequence, enhances prospects for retention of Earth-like or Venus-like atmospheres. However, where significant metals are lost in hydrogen escape, or the saturated phase lasts long enough to escape hydrogen and residual secondary components, then the mantle could be catastrophically depleted in volatiles, precluding later revival. 
\item We expect the sub-Earth and super-Earth regimes to present as a two-part cosmic shoreline. For rocky planets with escape velocities larger than Earth’s, the escape rate and threshold XUV flux are governed by the collisional-radiative thermostat, whereas for smaller escape velocities, the rates are energy-limited with a Knudsen onset. 
\end{enumerate}
Model mass losses for a range of planets spanning the cosmic shoreline are summarised in Table \ref{tab: equilibria}. The large XUV flux requirements for transonic escape could be relaxed in the case of significant Joule heating \citep{Cohen_2024}. Both sides of the cosmic shoreline are tentatively being populated \citep[e.g.,][]{Diamond-Lowe_2020, mansfield2024, Xue_2024, Cadieux_2024}. We hope the present study can aid in making population-level predictions for the retention of rocky exoplanet atmospheres. 
\section{Summary}
\label{sec: sum}
We conclude that the escape rate of secondary atmospheres from Mars-to-Earth-sized bodies can be roughly calculated with an energy limit above a threshold XUV flux given by the Knudsen-onset in collisionality at the sonic point. For super-Earths, escape shifts towards a transonic global ion outflow with the thermal structure determined by the balance of photoionization heating and forbidden line cooling. The transonic onset and escape rate can be roughly interpolated from the collisional-radiative thermostat presented in Figure \ref{fig: thermostat}. These results are consistent with the vulnerability of TRAPPIST-1 b and c to easily lose a thick atmosphere and the ability of LHS 1140 b to hold on to a possibly N$_2$-dominated secondary atmosphere.

We explored the landscape of high molecular weight escape (see Figure \ref{fig: conceit}) by varying the power-law index $\gamma$ corresponding to the steepness of atmospheric inversion driven by idealised XUV heating. Given the gravitational binding of the atmosphere described by escape parameter $\lambda_0$, analytic solutions to the atmospheric profiles were found, including approximations for the sonic-point properties and thermomechanical efficiency, which are relatively insensitive to a physically motivated range of lower boundary conditions. We compared our Young-Sun Earth transonic escape model with hydrocode simulations from \cite{Johnstone2019}.

We diagnosed forbidden-line cooling from polytropic profiles with a three-level atom model, finding it minor for Young-Sun Mars transonic escape and significant for Young-Sun Earth. We develop an explanation for the temperature behavior in varying XUV fluxes of \cite{Nakayama_2022} based on the feedback from electron densities on the forbidden line cooling in the photoionization front. We generalize the notion, calling it the collisional-radiative thermostat, and explore its control on escape under a range of conditions.

For strongly ionized atmospheres, the ambipolar electric field halves the mean molecular weight compared to a predominantly neutral flow, reducing the temperatures required for hydrostatic instability and transonic outflow by half. We suggest that incomplete accounting of these ion-electron interactions explains how \cite{Nakayama_2022} find persistent stability even up to $1000\times$ Modern Earth XUV fluxes, in contrast with our study, which finds transonic outflow for half that flux or less. We propose that the shift to ion-acoustic sound speed, characterized by the \textit{plasma} escape temperature, allows the transonic outflow of a secondary atmosphere from a super-Earth even under the action of the collisional-radiative thermostat (see Figure \ref{fig: thermostat}).

The work reported here marks the first 1D analytic modeling of the XUV-driven transonic escape of secondary atmospheres from rocky exoplanets. Our inclusion of both ion-electron effects and forbidden-line cooling  sets a precedent for modeling global loss of secondary atmospheres. Future studies involving state-of-the-art photochemistry, self-consistent hydrodynamics and molecular-kinetics simulations will be crucial in theoretically constraining Eta-Earth and the Cosmic Shoreline. \\

\software{Mathematica 14.1 \citep{Mathematica}} \\

R.D.C thanks Hamish Innes, Antonio Garc\'{\i}a Mu\~{n}oz, Brad Foley and Tad Komacek for providing feedback on the manuscript, Sarah Blumenthal for advice on photochemistry and Robert E. Johnson for interesting discussions. We thank the anonymous reviewer for comments that greatly improved the manuscript. This work was supported by the Science and Technology Facilities Council (STFC) and the Alfred P. Sloan Foundation under grant G202114194 (AEThER).

\bibliography{FP1}{}
\bibliographystyle{aasjournal}



\end{document}